\documentclass[11pt,nofootinbib]{article}
\pdfoutput=1
\usepackage{jheppub}
\usepackage{graphicx}
\usepackage{slashed}
\usepackage{subfig}
\usepackage[normalem]{ulem}
\usepackage{wasysym}
\usepackage{verbatim}
\usepackage{cancel}
\usepackage{natbib}
\usepackage{amsmath,amssymb,amsfonts}
\usepackage{enumerate}

\newcommand\beq{\begin{equation}}
\newcommand\eeq{\end{equation}}
\newcommand\Tr{\text{Tr}\,}

\newcommand\diag{\text{diag}}
\newcommand\hc{\text{h.c.}}

\title{Scalar phenomenology in type-II seesaw model}
\author[a]{R. Primulando,}
\author[b]{J. Julio}
\author[c]{and P. Uttayarat}

\affiliation[a]{Center for Theoretical Physics, Department of Physics, Parahyangan Catholic University, Jl. Ciumbuleuit 94, Bandung 40141, Indonesia}
\affiliation[b]{Indonesian Institute of Sciences (LIPI), Kompleks Puspiptek Serpong, Tangerang 15314, Indonesia}
\affiliation[c]{Department of Physics, Srinakharinwirot University, 114 Sukhumvit 23rd Rd., Wattana, Bangkok 10110, Thailand}

\emailAdd{rprimulando@unpar.ac.id}
\emailAdd{julio@lipi.go.id}
\emailAdd{patipan@g.swu.ac.th}

\abstract{ 
In this work we study the viable parameter space of the scalar sector in the type-II seesaw model. In identifying the allowed parameter space, we employ constraints from low energy precision measurements, theoretical considerations and the 125-GeV Higgs data. These tools prove effective in constraining the model parameter space. Moreover, the triplet also offers a rich collider phenomenology from having additional scalars that have unique collider signatures. We find that direct collider searches for these scalars can further probe various parts of the viable parameter space. These parts can be parametrized by the electroweak scalar triplet vacuum expectation value, the mass splitting of the singly- and doubly-charged scalars, and the doubly-charged Higgs mass. We find that different regions of the viable parameter space give rise to different collider signatures, such as the same-sign dilepton, the same-sign $W$ and the multilepton signatures. By investigating various LEP and LHC measurements, we derive the most updated constraints over the whole range of parameter space of the type-II seesaw model.   
}

\begin{document}
\maketitle

\flushbottom
\section{Introduction}

The Standard Model of particle physics (SM) has been very successful in describing various phenomena ranging across a wide range of energy scales that can be reached by current experiments. Despite this, the SM cannot be considered a complete theory because of its inability to generate neutrino masses. This is in direct contradiction with experimental results on neutrino oscillation, which have firmly established the massive, yet tiny, nature of the neutrinos. Hence the SM must be extended to accommodate small neutrino masses. The simplest but compelling way to do so is through the seesaw mechanism, where a dimension-5 operator is induced~\cite{Weinberg:1979sa}
\begin{equation}
	\frac{1}{\Lambda}L^iL^j\Phi^k\Phi^l\epsilon_{ik}\epsilon_{jl}.
\end{equation}
Here $L$ and $\Phi$ denote the lepton and Higgs doublets, respectively, while $i,j,k,l=1,$2 are $SU(2)_L$ indices, and $\epsilon\equiv i\sigma^2$ is an antisymmetric matrix. Due to the lepton-number-violating nature of this operator, the generated mass of the neutrino is of Majorana type, given by $m_\nu\sim v^2/(2\Lambda)$ with $v$ being the electroweak vacuum expectation value (vev). The smallness of the neutrino masses is due to the cutoff scale $\Lambda$. In the type-I seesaw mechanism, $\Lambda$ is  the mass of right-handed neutrinos transforming as singlets under the SM gauge groups \cite{Minkowski:1977sc,GellMann:1980vs,Yanagida:1979as,Mohapatra:1979ia}. Small neutrino masses of the order 0.1 eV, as implied by oscillation data, can be nicely explained by having a very large right-handed neutrino mass, i.e., about $10^{14}$ GeV. It is, however, very unlikely that this mechanism will be tested directly in collider experiments since it requires heavy states beyond reach of any ongoing or planned experiments.

Another variant of seesaw mechanism, the so-called type-II seesaw mechanism~\cite{Magg:1980ut,Schechter:1980gr,Cheng:1980qt,Lazarides:1980nt,Mohapatra:1980yp} also exists. Here, the SM is extended by a weak scalar triplet $\Delta$ that couples to a pair of lepton doublet via Yukawa interactions. Neutrino masses are induced after the triplet acquires a vev, i.e., $v_t\sim\mu v^2/M^2$ with $\mu$ and $M$ being the  coupling of trilinear term ($\mu \Phi\Delta^\dagger \Phi$) and the triplet mass, respectively. It is not hard to see that this relation possesses seesaw property as well. But in this model $\mu$ is allowed to be  small by the 't Hooft's naturalness principle. If this is the case, the new scale associated with triplet mass $M$ could be low. For instance, $\mu\sim v_t$ implies $M\sim v$, thereby making this model testable at colliders such as the Large Hadron Collider (LHC). 

The model with a weak-scale triplet is phenomenologically attractive because it can be produced at colliders through interactions with gauge bosons. For example, let us consider the doubly-charged component, $H^{++}$. This particle is mainly produced either in pair via $\gamma/Z$ exchanges, or in association with singly-charged partner $H^{-}$ via $W^+$ exchange. The cross sections of these two processes are quite large~\cite{Huitu:1996su,Muhlleitner:2003me}, so both the singly-charged and the doubly-charged scalars are produced in a large quantity at the LHC.\footnote{Photon-initianed production of doubly-charged scalars has been discussed in ref.~\cite{Babu:2016rcr}.} More importantly, the decays of $H^{++}$ could give rise to unique signatures that in turn can be used to constrain some parameter space of the model. Typical channels to look for are the same-sign dileptons (i.e., $H^{++}\to \ell^+\ell^+$), which are dominant in the regime of small triplet vev ($v_t\lesssim 10^{-4}$ GeV). Interestingly, branching ratios of these channels are dependent on neutrino mass matrix, and thus may help reveal the structure of neutrino mass matrix~\cite{Chun:2003ej,Garayoa:2007fw,Kadastik:2007yd,Akeroyd:2007zv,Perez:2008ha}. In fact, the CMS collaboration~\cite{CMS:2017pet}, invoking several benchmarks reflecting various neutrino mass patterns, have been able to exclude  the mass of $H^{++}$ lighter than 716-761 GeV. Stronger limit in the range of 800--850 GeV is obtained by both CMS and ATLAS by considering 100\% decay into light lepton pair ($ee$, $e\mu$ or $\mu\mu$)~\cite{Aaboud:2017qph,CMS:2017pet}. However, for larger $v_t$, the triplet Yukawa couplings become small since the product of the two is fixed by neutrino masses. As a result, the bounds derived above are not applicable because the dileptons are no longer the dominant decay channels. In fact for large $v_t$, the same-sign $W$ bosons decay becomes dominant. The importance of this channel for large $v_t$ has been discussed before in literature~\cite{Akeroyd:2009hb,Kanemura:2013vxa,Kanemura:2014goa,Kang:2014jia}. Last year, ATLAS could finally set the limit for this doubly-charged scalar decaying into same-sign on-shell $W$ bosons. In their analysis, ATLAS assumed the triplet vev $v_t=0.1$ GeV and the limit was found to be 220 GeV~\cite{Aaboud:2018qcu}, which is considerably weaker than those obtained from the dilepton channels. 

The bounds described above clearly do not apply to the entire parameter space of the model because they are derived from a set of simplifying assumptions. For instance, ATLAS assumed that the doubly-charged be produced through pair-production only, thus neglecting the other production mechanism, i.e., $H^{++}H^-$, which can be actually large.  CMS, on the other hand, did consider the $H^{++}H^-$ production in addition to the $H^{++}$ pair-production. Moreover,  ATLAS and CMS assumed the degeneracy in triplet masses, so the decays of $H^{++}$ are limited to the same-sign dilepton and the same-sign diboson channels.

In a more general approach, the triplet masses should be allowed to split so that the triplet can undergo cascade decays, e.g., $H^{++} \to H^+W^{+*} \to \phi W^{+*}W^{+*}$ with $\phi$ being the neutral  components. This scenario has been realized in previous works~\cite{Chakrabarti:1998qy,Chun:2003ej,Akeroyd:2005gt,Melfo:2011nx,Aoki:2011pz,Akeroyd:2011zza,Chiang:2012dk,Chun:2012zu,Chun:2013vma,Han:2015hba,Han:2015sca,Akeroyd:2012nd,delAguila:2008cj,Mitra:2016wpr,Du:2018eaw,Antusch:2018svb}. Many of these works conclude that in the case of nondegenerate scalar masses, the cascade decays are important in constraining the model parameter space. The interesting thing about cascade decays is, as particularly pointed out in ref.~\cite{Mitra:2016wpr}, they can lead to multilepton  signatures. As CMS data exist~\cite{Sirunyan:2017lae}, to the best of our knowledge,  these signatures have never been used to constrain the parameter space of the type II seesaw model.


In this paper, we perform a collider analysis to find an up-to-date collider bounds. In order to get a more realistic picture, we consider all relevant production channels. In addition, the three triplet scalar masses are allowed to split. We then utilize the CMS multilepton analysis \cite{Sirunyan:2017lae} to set bound on the scalar masses. Such multilepton analysis proves effective in constraining the model parameter space to such an extent that it  can probe regions not covered by the current official LHC searches. Previous analyses emphasizing the multilepton search have been discussed in~\cite{Chun:2012zu,Chun:2013vma,Han:2015hba,Han:2015sca,Akeroyd:2012nd,delAguila:2008cj,Mitra:2016wpr,Du:2018eaw,Antusch:2018svb}. To check the self-consistency of the model we also impose several constraints that can significantly affect the parameter space of the model.  Among these are the electroweak oblique parameters and the Higgs data. It is well known that the triplet can affect the values of the oblique parameters, in particular the $S$ and $T$ parameters~\cite{Lavoura:1993nq,Chun:2012jw}.  The presence of triplet will affect the properties of the 125-GeV Higgs boson ($h$) that have been measured to about $10\%$ accuracy~\cite{Khachatryan:2016vau}. In particular, the two charged components can alter significantly the partial decay width of the loop-induced $h \rightarrow \gamma\gamma$~\cite{Arhrib:2011vc,Akeroyd:2012ms,Dev:2013ff,Chun:2012jw,Das:2016bir}.  Moreover, the triplet Yukawa couplings may induce lepton flavor violations at tree level, especially when the triplet is light. We combine all these precision constraints with the current LHC constraints to understand the viable parameter space of the model as well. 

This paper is organized as follows. In section~\ref{sec:model}, we set up our notation and convention. We identify the viable parameter space of the model in section~\ref{sec:const}. The LHC phenomenology of the model is discussed in section~\ref{sec:ColliderBounds}. We then conclude in section~\ref{sec:conclusion}. 

\section{The model}
\label{sec:model}
In this section, we give a brief overview of the type-II seesaw model. Here we demonstrate how scalar masses are expressed in terms of model parameters. It is followed by a discussion on neutrino mass generation in the framework of type-II seesaw mechanism.

\subsection{Scalar sector}
\label{sec:scalar}
In the type-II seesaw model, the scalar sector is extended by adding a weak scalar triplet, in addition to the usual scalar doublet. Explicitly, one can write
\begin{align}
\Phi(1,2,+1/2)=\begin{pmatrix}  \phi^+ \\ \phi^0   \end{pmatrix} \quad {\rm and} \quad \Delta(1,3,+1)=\begin{pmatrix}  \Delta^+/\sqrt{2} & \Delta^{++} \\ \Delta^0 & -\Delta^+/\sqrt{2}  \end{pmatrix}.
\label{sc}
\end{align}
The numbers in parentheses denote their representations under $SU(3)_C \times SU(2)_L \times U(1)_Y$ gauge groups of the SM.
The kinetic terms, containing interactions between scalars and gauge bosons, are given by
\begin{equation}
	\mathcal{L}_{kin} \supset (D^\mu\Phi)^\dagger(D_\mu\Phi)+\Tr\left[(D^\mu\Delta)^\dagger(D_\mu\Delta)\right],
\end{equation}
with covariant derivatives
\begin{align}
	D_\mu\Phi &= \partial_\mu\Phi - i\frac{g}{2}W^a_\mu\sigma^a\Phi - i\frac{g'}{2}B_\mu\Phi,\\
	D_\mu\Delta &= \partial_\mu\Delta - i\frac{g}{2}\left[W^a_\mu\sigma^a,\Delta\right] -ig'B_\mu\Delta.
\end{align}
Here $\sigma^a$ with $a=1,2,3$ are Pauli matrices, whereas $g$ and $g'$ are the $SU(2)_L$ and $U(1)_Y$ gauge couplings, respectively. The most general renormalizable scalar potential is given by\footnote{Other conventions for the scalar potential exist in the literature. All of them can be written in terms of the others, see for example \cite{Bonilla:2015eha}.}
\begin{align}
	V(\Phi,\Delta) &= -m_\Phi^2\Phi^\dagger\Phi+M^2\Tr\Delta^\dagger\Delta + (\mu\,\Phi^Ti\sigma^2\Delta^\dagger\Phi + {\rm h.c.}) + \frac{\lambda}{4}(\Phi^\dagger\Phi)^2 \nonumber \\
	&\quad + \lambda_1\Phi^\dagger\Phi\Tr \Delta^\dagger\Delta +\lambda_2(\Tr\Delta^\dagger\Delta)^2 + \lambda_3\Tr(\Delta^\dagger\Delta)^2 + \lambda_4\Phi^\dagger\Delta\Delta^\dagger\Phi.
\label{sc-pot}
\end{align}
Without loss of generality, all mass-squared parameters as well as quartic couplings appearing in eq.~(\ref{sc-pot}) can be taken to be real. Moreover, the trilinear coupling $\mu$ can be taken to be positive by absorbing its phase into $\Phi$ or $\Delta$. 

After spontaneous symmetry  breaking, both scalars $\Phi$ and $\Delta$ acquire vevs, denoted as $\left<\Phi \right> =v_d/\sqrt{2}$ and $\left<\Delta \right> =v_t/\sqrt{2}$, through the minimization of the scalar potential. The two minimization conditions, from which we can express $m_\Phi^2$ and $M^2$ in terms of other parameters, are 
\begin{align}
	m_\Phi^2 &= \frac{\lambda}{4}v_d^2 + \frac{\lambda_1+\lambda_4}{2}v_t^2 - \sqrt{2}\mu v_t, \\
	M^2 &= -\frac{\lambda_1+\lambda_4}{2}v_d^2 - (\lambda_2+\lambda_3)v_t^2 + \frac{\mu v_d^2}{\sqrt{2}v_t}.
	\label{min-conds}
\end{align}
We see that the presence of the $\mu$ term can induce triplet vev without even requiring $M^2<0$. The two vevs will contribute to  masses of  $W^\pm$ and $Z$ gauge bosons, which are $m_W^2=g^2(v_d^2+2v_t^2)/4$ and $m_Z^2=(g^2+g'^2)(v_d^2+4v_t^2)/4$, and thus the electroweak vev can be inferred as $v^2 \equiv v_d^2+2v_t^2= (246~{\rm GeV})^2$. The ratio of these two gauge boson masses is constrained by the $\rho$ parameter, defined as $\rho\equiv m_W^2/(m_Z^2\cos^2\theta_W)$. In the SM at tree level we have $\rho=1$, which is in perfect agreement with the observed value from electroweak precision measurements that set $\rho_{\rm obs.} = 1.00039 \pm 0.00019$.  In this model, such relation is not trivially satisfied because 
\begin{align}
\rho = 1+\Delta\rho = 1-\frac{2v_t^2}{v_d^2+4v_t^2},
\end{align}
Therefore, complying with the data implies an upper bound on $v_t$, namely, $v_t\leq 4.8$ GeV at 95\% CL.

Throughout this paper we assume CP invariance in the scalar sector, and hence all vevs are  real. As a consequence, there will be no mixing between the real and the imaginary parts of the scalars. In more explicit form we write both scalars as
\begin{equation}
	\Phi = \begin{pmatrix} \phi^+\\ (v_d+h_d+iz_d)/\sqrt{2}\end{pmatrix},\quad
	\Delta = \begin{pmatrix} \Delta^+/\sqrt{2} & \Delta^{++}\\ (v_t + h_t + iz_t)/\sqrt{2} & ~~-\Delta^+/\sqrt{2}\end{pmatrix}.
\end{equation}
Out of these 10 degrees of freedom, three become Goldstone bosons, eaten up by $W$ and $Z$ bosons during the electroweak symmetry breaking. What remain are seven fields, consisting of doubly-charged scalars $H^{\pm\pm}\equiv\Delta^{\pm\pm}$, singly-charged scalars $H^{\pm}$, CP-even scalars $H$ and $h$, and CP-odd scalar $A$. The doubly-charged Higgs comes solely from the triplet with mass given by
\begin{align}
m_{H^{++}}^2 = \frac{\mu v_d^2}{\sqrt{2}v_t} - \frac{\lambda_4}{2}v_d^2 - \lambda_3v_t^2. 
\label{dch}
\end{align}

The other sectors involve mixing. For instance, the mixing between $\phi^+$ and $\Delta^+$ is characterized by mass matrix 
\begin{align}
M_{\rm charged}^2 = \left(
\begin{array}{cc}
 \sqrt{2} \mu  v_t-\frac{\lambda_4 v_t^2}{2} & -\mu v_d + \frac{\sqrt{2}}{4} \lambda_4 v_tv_d \\
 -\mu v_d + \frac{\sqrt{2}}{4} \lambda_4 v_tv_d & \frac{\mu v_d^2 }{\sqrt{2}v_t}-\lambda_4 v_d^2 \\
\end{array}
\right).
\label{charged}
\end{align}
One can inspect that the mass matrix in eq.~(\ref{charged}) has zero determinant but nonzero trace, meaning that one of its two eigenvalues is zero, while the other is not. The spectrum with vanishing eigenvalue is to be identified as the Goldstone boson eaten up by $W^+$ boson, and it is defined as $G^+ \equiv \cos\beta\, \phi^+ + \sin\beta\, \Delta^+$. The other combination $H^+ \equiv -\sin\beta\, \phi^+ + \cos\beta\,\Delta^+$, orthogonal to $G^+$,  is the physical field with mass
\begin{align}
m^2_{H^+} = \frac{(2\sqrt{2}\mu-\lambda_4v_t)}{4v_t}(v_d^2+2v_t^2).
\label{sch}
\end{align}
The mixing angle is found to be $\tan\beta=\sqrt{2}v_t/v_d$, which is always small due to the constraint on $v_t$ from the $\rho$ parameter. This shows that the singly-charged particle is predominantly triplet.

For the neutral sector, since we assume CP invariance in scalar sector, the mixing can only occur between states with the same CP property. In the CP-odd case, the imaginary parts of neutral $\Phi$ and $\Delta$  mix through
\begin{align}
M^2_{\rm odd} = \begin{pmatrix} 
 2 \sqrt{2} \mu v_t & -\sqrt{2} \mu v_d  \\
 -\sqrt{2} \mu v_d  & \frac{\mu v_d^2 }{\sqrt{2} v_t}
\end{pmatrix}.
\end{align}
As in the singly-charged case, this mass matrix also has zero determinant and nonzero trace. The massless field, $G^0\equiv\cos\beta'\,z_d+\sin\beta'\,z_t$, is the Goldstone boson eaten up by the $Z$ gauge boson, while the massive one is the physical CP-odd scalar, $A\equiv-\sin\beta'\,z_d+\cos\beta'\,z_t$, with mass 
\begin{align}
m_A^2 = \frac{\mu}{\sqrt{2}v_t}(v_d^2+4v_t^2).
\label{mA}
\end{align}
The mixing angle in this case is found to be small as well, i.e.,  $\tan\beta'=2v_t/v_d$, meaning that the CP-odd particle is predominantly triplet.

For the CP-even scalars the mass matrix of $h_d$ and $h_t$ is given by
\begin{align}
M_{\rm even}^2 = \begin{pmatrix}
\tfrac{\lambda}{2}v_d^2  & -\sqrt{2}\mu v_d + (\lambda_1+\lambda_4)v_tv_d \\ -\sqrt{2}\mu v_d + (\lambda_1+\lambda_4)v_tv_d  & \frac{\mu v_d^2}{\sqrt{2}v_t} + 2(\lambda_2+\lambda_3)v_t^2
\end{pmatrix}.
\label{even}
\end{align}
Similar to what we did previously, the mass matrix of eq.~(\ref{even}) can be brought into its diagonal form by rotating $h_d$ and $h_t$ into physical eigenstates, $h\equiv \cos\alpha\, h_d + \sin\alpha\, h_t$ and $H\equiv -\sin\alpha\, h_d+\cos\alpha\, h_t$. From here, the masses of these CP-even scalars are determined through
\begin{align}
\begin{pmatrix} m_h^2 & 0 \\ 0 & m_H^2 \end{pmatrix} = \left(\begin{array}{cc} \cos\alpha & \sin\alpha \\ -\sin\alpha & \cos\alpha \end{array}\right)
M^2_{\rm even} \begin{pmatrix} \cos\alpha & -\sin\alpha \\ \sin\alpha & \cos\alpha \end{pmatrix}
\label{higgs-mass}
\end{align}
with mixing angle
\begin{align}
\tan2\alpha = \frac{-2\sqrt{2}\mu v_d + 2(\lambda_1+\lambda_4)v_tv_d}{\frac{\lambda}{2}v_d^2 -  \frac{\mu v_d^2}{\sqrt{2}v_t} - 2(\lambda_2+\lambda_3)v_t^2}.\label{angle-even}
\end{align}
In contrast to the case of singly-charged and CP-odd Higgs, wherein their mixing angles (i.e., $\beta$ and $\beta'$) are naturally small, here the mixing angle $\alpha$ is allowed to  be large, in particular when the diagonal entries of $M^2_{\rm even}$ are exactly the same, leading to the cancellation of the denominator of eq.~(\ref{angle-even}), hence  maximal mixing. As a result the two masses, $m_h$ and $m_H$, become nearly degenerate.  (Exact degeneracy requires delicate cancellation in off-diagonal entries as well, implying $\alpha=0$.) However this maximal-mixing scenario seems to be disfavored  by recent LHC Higgs data, discussed in more detail in section~\ref{sec:viable}, constraining $|\sin\alpha|\lesssim0.3$ at 95\% CL.\footnote{The phenomenology of the large mixing in the CP-even neutral case is considered in refs. \cite{Akeroyd:2010je, Dey:2008jm}.} Taking this bound into account, it is then reasonable to consider that $h$ be predominantly doublet. Hence, we identify it as the 125-GeV resonance observed at the LHC. One should note, however, that $h$ could be the lighter or the heavier of the two eigenstates.

\subsection{Neutrino sector}
As for neutrino mass generation, the triplet within this model is known for its ability to induce Majorana neutrino masses  via interactions with the left-handed lepton doublet $L\equiv(\nu,\ell)^T$
\begin{align}
{\cal L}_{Y}^{\rm new} = f_{ab} L^T_a C i\sigma^2 \Delta L_b + \hc
\label{yuk}
\end{align}
Here $C$ is the charge conjugation matrix and $a,b=e,\mu,\tau$ are flavor indices. The Yukawa couplings $f_{ab}$ are  symmetric under $a \leftrightarrow b$ interchange, thanks to Fermi statistics and the $SU(2)_L$ contraction property. 
Expanding eq.~(\ref{yuk}) in terms of triplet components yields
\begin{align}\mathcal{L}^{\rm new}_{Y} =& -f_{ab} \bar{\ell^c_a}\ell_{b}\Delta^{++} - \sqrt{2}f_{ab} \bar{\nu^c_a} \ell_{b} \Delta^+ + f_{ab} \bar{\nu^c_a}  \nu_{b} \Delta^0 + \hc,
\label{yuk-exp}
\end{align}
where $\psi^c$ denotes right-handed antilepton field.

Once the triplet develops a vev, the last term of eq.~(\ref{yuk-exp}) will induce a Majorana-type mass matrix, i.e., $M_\nu =2f\left<\Delta\right> = \sqrt{2} f v_t$. 
The matrix $M_\nu$ can be expressed in terms of neutrino oscillation parameters via
\begin{align}
M_\nu = U^* \begin{pmatrix} m_1 & 0 & 0 \\ 0 & m_2 & 0 \\ 0 & 0 & m_3 \end{pmatrix} U^\dagger,
\label{nu-mass}
\end{align}
where we use the convention that neutrino mass eigenvalues are real, thereby making the two Majorana phases, $\alpha_1$ and $\alpha_2$,  appear in the Pontecorvo-Maki-Nakagawa-Sakata (PMNS) matrix, $U$, parametrized as \cite{Tanabashi:2018oca} 
\begin{align}
U = \begin{pmatrix} 1 & 0 & 0 \\ 0 & c_{23} & s_{23} \\ 0 & -s_{23} & c_{23} \end{pmatrix}
\begin{pmatrix} c_{13} & 0 & s_{13}e^{-i\delta} \\ 0 & 1 & 0 \\ -s_{13}e^{i\delta} & 0 & c_{13} \end{pmatrix}
\begin{pmatrix} c_{12} & s_{12} & 0 \\ -s_{12} & c_{12} & 0 \\ 0 & 0 & 1 \end{pmatrix}
\times
\diag(1, e^{i\alpha_1/2}, e^{i\alpha_2/2}).
\end{align}
Here $s_{ij}$ ($c_{ij}$) denotes the sine (cosine) of the mixing angle $\theta_{ij}$, and $\delta$ denotes the Dirac CP phase, the only phase contributing to neutrino oscillations. 

For our analysis, we will use the global fit of neutrino data performed  by the NuFIT Group~\cite{Esteban:2018azc}, summarized in table~\ref{nu-fit}. (For similar analysis, see also ref.~\cite{deSalas:2017kay}.)  Some best-fit values of neutrino oscillation parameters are indeed showing dependence on mass hierarchy, and for that reason the global fit values are given for each hierarchy. The model itself can admit both the normal hierarchy (NH) and the inverted  hierarchy (IH) of neutrino masses. 

\begin{table}
\begin{center}
\begin{tabular}{|c|c|c|}
\hline\hline 
Parameter & Normal Hierarchy & Inverted Hierarchy \\
\hline 
$\sin^2\theta_{12}$ & $0.310^{+0.013}_{-0.012}$ & $0.310^{+0.013}_{-0.012}$ \\
$\sin^2\theta_{23}$ & $0.582^{+0.015}_{-0.019}$ & $0.582^{+0.015}_{-0.018}$ \\
$\sin^2\theta_{13}$ & $0.02240^{+0.00065}_{-0.00066}$ & $0.02263^{+0.00065}_{-0.00066}$ \\
$\delta~[^\circ]$ & $217^{+40}_{-28}$ & $280^{+25}_{-28}$ \\
$\Delta m^2_{21}~[10^{-5}~{\rm eV}^2]$ & $7.39^{+0.21}_{-0.20}$ & $7.39^{+0.21}_{-0.20}$ \\
$\Delta m^2_{\rm atm.}~[10^{-3}~{\rm eV}^2]$ & $2.525^{+0.033}_{-0.031}$ & $-2.512^{+0.034}_{-0.031}$ \\
\hline
\hline
\end{tabular}
\caption{The global fit of neutrino oscillation parameters. We use the convention of ref.~\cite{Esteban:2018azc}, i.e., $\Delta m^2_{\rm atm.} \equiv \Delta m^2_{31}$ in Normal Hierarchy case and $\Delta m^2_{\rm atm.} \equiv \Delta m^2_{32}$ in Inverted Hierarchy case.}
\label{nu-fit}
\end{center}
\end{table}

The Dirac CP phase $\delta$ is of particular interest because of its ability to induce the leptonic CP violation. Besides, the latest results from T2K \cite{Abe:2017vif,Abe:2018wpn} and NOvA \cite{Adamson:2017gxd,NOvA:2018gge} experiments suggest that such angle differs from zero. Furthermore, combining them into global fit of neutrino data contributes to excluding even $\delta=\pi/2$ by about $3\sigma$ \cite{Esteban:2018azc,deSalas:2017kay}. This result is reflected in its best-fit value found to be $\delta=217^\circ$ ($280^\circ$) for the NH (IH) case. However, it is worth noting that the $2\sigma$ range of $\delta$ in the NH case is very close to $180^\circ$, so it cannot yet be considered as robust evidence for CP violation. In spite of this we still use the best-fit value of $\delta$ in our calculation.  On the contrary, Majorana phases remain hitherto unconstrained and can take any value in the range of $0\leq \alpha_1,\alpha_2\leq 2\pi$. In most of our analysis, unless stated otherwise, we will simply set them to zero.

For the purpose of deriving collider bounds, we will consider the case of normal hierarchy with the lightest neutrino being massless ($m_1=0$). However, for completeness, in the Appendix we will also consider the case of inverted hierarchy with $m_3 = 0$. Additionally, the case where all neutrino masses are nearly degenerate with the lightest neutrino taking 0.1 eV of mass is also studied in the appendix. Using eq.~(\ref{nu-mass}) with parameter best-fit values taken from table~\ref{nu-fit}, the explicit form of $M_\nu$, in the unit of $10^{-3}$ eV, for each case is given as follows:
\begin{enumerate}[(i)]
\item {\it Normal hierarchy}
\begin{align}
& M_\nu = \left(
\begin{array}{rlrlrl}
 3.11& e^{i0.355} & \,\,\, 4.00 &\,\,\, e^{-i2.02} & 7.30 &\,\,\, e^{-i2.71} \\
 4.00&e^{-i2.02} & \,\,\, 31.6&\,\,\, e^{-i0.0123} & 21.2 &\,\,\, e^{i0.00143} \\
 7.30& e^{-i2.71} &\,\,\, 21.2&\,\,\, e^{i0.00143} & 23.5 &\,\,\, e^{i0.0140} \\
\end{array}
\right) \quad {\rm with}~m_1=0,  \nonumber \\
~~  \label{numass-nh}
\\[-1em]
& M_\nu = \left(
\begin{array}{rlrlrl}
 98.6&e^{i0.0244} & \,\,\, 14.4&e^{-i1.64} & \,\,\,12.3&e^{-i1.65} \\
 14.4&e^{-i1.64} & \,\,\,106&e^{-i0.0120} & \,\,\,4.93&e^{-i0.22} \\
 12.3&e^{-i1.65} & \,\,\, 4.93&e^{-i0.22} & \,\,\,104&e^{-i0.0085} \\
\end{array}
\right) \quad {\rm with}~m_1=0.1~{\rm eV}. \nonumber
\end{align}

\item {\it Inverted hierarchy }
\begin{align}
& M_\nu = \left(
\begin{array}{rlrlrl}
 48.5& &\,\,\, 5.59&e^{-i1.71} &\,\,\, 4.82&e^{-i1.80} \\
 5.59&e^{-i1.71} &\,\,\, 20.2&e^{i0.00856} &\,\,\, 25.1&e^{i3.13} \\
 4.82&e^{-i1.80} &\,\,\, 25.1&e^{i3.13} & \,\,\,28.6&e^{i0.00737} \\
\end{array}
\right) \quad\quad {\rm with}~m_3=0, \nonumber \\
~~ \label{numass-ih}\\[-1em]
& M_\nu = \left(
\begin{array}{rlrlrl}
 107&e^{-i0.00723} &\,\,\, 23.6&e^{-i1.58} &\,\,\, 20.0&e^{-i1.59} \\
 23.6&e^{-i1.58} &\,\,\, 102&e^{i0.00470} &\,\,\, 8.10&e^{i3.09} \\
 20.0&e^{-i1.59} &\,\,\, 8.10&e^{i3.09} &\,\,\, 105&e^{i0.0037} \\
\end{array}
\right)\quad {\rm with}~m_3=0.1~{\rm eV}. \nonumber
\end{align}
\end{enumerate}

Given that $M_\nu$ is fixed by neutrino oscillation data, the magnitude of couplings $f_{ab}$ are dependent on $v_t$. For example, extremely small triplet vev like $v_t \sim 0.1$ eV implies $f_{ab} \sim 1$, which are large enough to play a significant role in same-sign dilepton decay channels $H^{++}\to \ell^+_a\ell^+_b$. In fact,  ATLAS collaboration \cite{Aaboud:2017qph} recently have excluded this kind of doubly-charged scalar  lighter than about 850 GeV by assuming 100\% decay into same-sign muons (albeit not truly reflecting the structure of neutrino mass matrix). In the type-II seesaw context, this exclusion only applies when $v_t$ is small (i.e., $v_t \lesssim10^{-4}$ GeV) so that other decay channels whose couplings are proportional to $v_t$, like $H^{++} \to W^+W^+$, are suppressed. Of course, when $v_t$ is greater than $10^{-4}$ GeV, or triplet masses are allowed to split, the phenomenology becomes much richer because other decay channels become relevant. We will discuss this in more details in section~\ref{sec:ColliderBounds}. 

Finally we note that a triplet with mass around a few hundreds GeV will induce tree-level lepton flavor violation (LFV) processes, e.g., $\mu \to 3e$, especially when $v_t\lesssim 10^{-7}$ GeV where the Yukawa couplings are sizable. In this case, bounds derived from the LFV processes can be so strong that they may overpower collider bounds.  We will discuss the LFV bounds in more details in section~\ref{sec:const}.

\allowdisplaybreaks
\section{Constraints on model parameters}
\label{sec:const}
In this section, we will elaborate more on the allowed parameter space discussed briefly in the previous section. In order to get a better understanding of such parameter space, it is helpful to characterize the parameter space in terms of physical masses, which can be directly checked against results from LHC searches. All five quartic couplings and the trilinear coupling, $\mu$, can be expressed in terms of five scalar masses and mixing angle $\alpha$ by using eqs.~(\ref{dch}), (\ref{sch}), (\ref{mA}) and (\ref{higgs-mass}), as done in \cite{Arhrib:2011uy}. However, one should note that both $\lambda_2$ and $\lambda_3$ always appear with $v_t$ in all expressions of scalar masses. Thus, we expect their impacts are negligible compared to other terms, even when they take on the largest possible values allowed by perturbativity.\footnote{In this work, a scalar quartic coupling is considered perturbative if its magnitude is less than $4\pi$.} It is more practical to write, instead, 
\begin{align}
\lambda  = & \frac{2m_h^2}{v^2-2v_t^2} + \left[\frac{4m_{H^+}^2-2m_{H^{++}}^2-2m_h^2+(-8m_{H^+}^2+(4\lambda_2+2\lambda_3)v^2)(v_t/v)^2}{v^2-2v_t^2}\right]t_\alpha^2, \\
\lambda_1 =  &\frac{2(m_{H^{++}}^2+\lambda_3v_t^2)}{v^2-2v_t^2} + \left[\frac{m_h^2-2m_{H^+}^2+m_{H^{++}}^2+(4m_{H^{+}}^2-(2\lambda_2+\lambda_3)v^2)(v_t/v)^2}{ v^2-2v_t^2}\right] \nonumber \\ & \qquad\qquad \qquad\qquad \times \frac{\sqrt{v^2-2v_t^2}}{v_t}t_\alpha, \\
\lambda_4 = & \frac{4(m_{H^+}^2-m_{H^{++}}^2) - (8m_{H^+}^2+4\lambda_3 v^2)(v_t/v)^2}{v^2-2v_t^2}, \\
 m_A^2  = & \left[\frac{2m_{H^+}^2-m_{H^{++}}^2}{v^2-2v_t^2} - \frac{(4m_{H^+}^2+\lambda_3 v^2)(v_t/v)^2}{v^2-2v_t^2}\right] (v^2+2v_t^2), \\
m_H^2  = & -m_h^2t_\alpha^2 + \left[2m_{H^+}^2-m_{H^{++}}^2 - (4m_{H^+}^2-(2\lambda_2+\lambda_3)v^2)(v_t/v)^2\right](1+t_\alpha^2),
\end{align} 
where $t_\alpha\equiv\tan\alpha$, and we have used the relation $v=\sqrt{v_d^2+2v_t^2}$. As expected, all contributions of $\lambda_2$ and $\lambda_3$ come with suppression factor $v_t/v$. It is then instructive to expand the above expressions in power series of $v_t/v$, with the lowest order terms found to be 
\begin{align}
\lambda \simeq& \frac{2m_h^2}{v^2}(1-t_\alpha^2) + \left[\frac{4m_{H^+}^2-2m_{H^{++}}^2}{v^2} \right]t_\alpha^2 
\label{lambda} \\
 \lambda_1 \simeq& \frac{2m_{H^{++}}^2}{v^2} + \left[\frac{m_h^2-2m_{H^+}^2+m_{H^{++}}^2}{v^2} \right]\left(\frac{v}{v_t}\right)t_\alpha \label{eq:lambda1}\\
 \lambda_4 \simeq& \frac{4(m_{H^+}^2-m_{H^{++}}^2)}{v^2} \label{eq:lambda4} \\
 m_A^2 \simeq & ~2m_{H^+}^2-m_{H^{++}}^2 \label{eq:mA}\\
 m_H^2 \simeq& ~ (2m_{H^+}^2-m_{H^{++}}^2)(1+t_\alpha^2) - m_h^2t_\alpha^2
 \label{mH}
\end{align}
It is interesting to see that, for sufficiently small $\sin\alpha$, $m_A$ and $m_H$ become nearly degenerate, and the doublet self-quartic coupling $\lambda$ is close to the SM value, i.e., $2m_h^2/v^2=0.516$. Note that eq. (3.9) implies the triplet masses-squared are almost equally split, i.e., $m_{H^+}^2 - m_{H^{++}}^2 \simeq m_A^2 - m_{H^+}^2 \simeq \lambda_4v^2/4$. With this relation in hand, we can present the allowed parameter space of the model in the $m_{H^{++}}-\Delta m$ plane where $\Delta m\equiv m_{H^+}-m_{H^{++}}$.

The advantage of writing couplings and masses as in eqs. (\ref{lambda})-(\ref{mH}) is that we can easily check the compatibility of these quartic couplings against various constraints, in particular $h\to\gamma\gamma$ and vacuum stability conditions of the scalar potential, by which the model allowed parameter space can be derived.  Orthogonal bounds, like the ones coming from LFV and collider, will be discussed separately. Nevertheless, they will be overlaid on the allowed parameter space obtained from the scalar sector so that we can get a complete picture of the model.

\subsection{Perturbativity, vacuum stability and perturbative unitarity constraints}
\label{sec:theoretical}
All quartic couplings appearing in scalar potential (\ref{sc-pot}) must be such  that the scalar potential remains bounded from below in any directions of field space. This, known as vacuum stability, requires the following conditions to be satisfied simultaneously \cite{Bonilla:2015eha,Arhrib:2011uy}\footnote{An alternative derivation for different notation of scalar potential has also been given in ref.~\cite{Babu:2016gpg}.}
\begin{align}
&({\rm i})~ \lambda \geq0, \quad ({\rm ii})~ \lambda_2+\frac{\lambda_3}{2}\geq0, \quad ({\rm iii})~\lambda_1+\sqrt{\lambda(\lambda_2+\lambda_3)} \geq 0, \quad ({\rm iv})~\lambda_1+\lambda_4+\sqrt{\lambda(\lambda_2+\lambda_3)}\geq0, \nonumber \\
&({\rm v})~|\lambda_4|\sqrt{\lambda_2+\lambda_3}-\lambda_3\sqrt{\lambda} \geq0 \quad {\rm or} \quad 
2\lambda_1+\lambda_4+\sqrt{(2\lambda-\lambda_4^2/\lambda_3)\left(2\lambda_2+\lambda_3\right)}\geq0.
\label{eq:stability}
\end{align}
Note that conditions (i), (ii) and (v) are easy to satisfy. Condition (i), for example, always holds true because $\lambda\simeq0.516$ for most cases. As for conditions (ii) and (v), they can be simultaneously satisfied by choosing appropriate values of $\lambda_2$ and $\lambda_3$, which have negligible impacts on Higgs data.

Quartic couplings of eq.~(\ref{sc-pot}) can also contribute to  tree-level two-body scatterings, whose amplitudes are required to be unitary in all orders of perturbation calculation~\cite{Cornwall:1974km,Dicus:1992vj}. 
In the present model, perturbative unitarity will be preserved provided~\cite{Arhrib:2011uy}
\begin{align} 
&	\left|(\lambda+4\lambda_2+8\lambda_3) \pm \sqrt{(\lambda-4\lambda_2-8\lambda_3)^2+16\lambda_4^2}\right| \le64 \pi,\\
&	\left|(3\lambda+16\lambda_2+12\lambda_3) \pm \sqrt{(3\lambda-16\lambda_2-12\lambda_3)^2+24(2\lambda_1+\lambda_4)^2}\right| \le 64\pi, \label{pert-uni} \\
&|\lambda|\leq 32\pi, \\
&|2\lambda_1+3\lambda_4|\leq 32\pi, \\
&|2\lambda_1-\lambda_4| \leq 32\pi ,\\
&|\lambda_1|\leq 16\pi, \\
&|\lambda_1+\lambda_4| \leq 16\pi, \\
&|2\lambda_2-\lambda_3| \leq 16\pi, \\
&|\lambda_2| \leq 8\pi, \\
&|\lambda_2+\lambda_3| \leq 8\pi.
\end{align}
It turns out that out of the 10 conditions given above, only (\ref{pert-uni}) gives a nontrivial constraint. This condition sets an upper bound on $\lambda_2+\lambda_3$, which is instrumental in deriving the allowed parameter space of the model. 
Other relations are automatically satisfied as long as the couplings are perturbative.

\subsection{Electroweak precision constraints}
\label{sec:ewpc}
We have seen before that the electroweak $\rho$ parameter places an upper limit on $v_t$, i.e., $v_t\le4.8$ GeV at 95\% CL.
To obtain further constraints on the model parameter space, we consider the electroweak oblique parameters. We will focus on the $S$ and $T$ parameters. They are given by~\cite{Chun:2012jw}
\begin{equation}
\begin{aligned}
	S &= -\frac{1}{3\pi}\ln\frac{m_{H^{++}}^2}{m_H^2} - \frac{2}{\pi}\left[(1-2s_W^2)^2\xi\left(\frac{m_{H^{++}}^2}{m_Z^2}\right) + s_W^4\xi\left(\frac{m_{H^{+}}^2}{m_Z^2}\right) + \xi\left(\frac{m_{H}^2}{m_Z^2}\right)\right],\\
	T &= \frac{1}{8\pi c_w^2s_w^2}\left[\eta\left(\frac{m_{H^{++}}^2}{m_Z^2},\frac{m_{H^+}^2}{m_Z^2}\right) + \eta\left(\frac{m_{H^+}^2}{m_Z^2},\frac{m_{H}^2}{m_Z^2}\right)\right],
\end{aligned}
\label{eq:SandT}
\end{equation}
where
\begin{equation}
\begin{aligned}
	\xi(x) &= \frac49 - \frac43x + \frac{1}{12}(4x-1)f(x),\\
	f(x) &= \left\{\begin{aligned}&-4\sqrt{4x-1}\arctan \frac{1}{\sqrt{4x-1}},\quad \text{for }4x>1,\\
		&\sqrt{1-4x}\ln\frac{2x-1+\sqrt{1-4x}}{2x-1-\sqrt{1-4x}},\quad \text{for }4x\le1,\end{aligned}\right.\\
	\eta(x,y) &= x + y - \frac{2xy}{x-y}\ln\frac{x}{y}.
\end{aligned}
\end{equation}
For our numerical analysis, we take the best fitted value for $S$ and $T$ parameters provided by the Gfitter Group: $S=0.06\pm0.09$ and $T=0.10\pm0.07$ with a correlation coefficient +0.91~\cite{Baak:2014ora}.

\subsection{Higgs data}
\label{sec:higgsfit}

The couplings of the 125-GeV Higgs boson has been studied extensively at the LHC. These Higgs data provide a nontrivial constraint on the type-II seesaw parameter space. These couplings can be parametrized by~\cite{Carmi:2012in,Espinosa:2012im}
\begin{equation}
\begin{split}
	\mathcal{L}_h &\supset  c_W\frac{2m_W^2}{v}hW^+_\mu W^{-\mu} + c_Z\frac{m_Z^2}{v}hZ_\mu Z^{\mu} - c_f\frac{m_f}{v}h\bar f f \\
	&\qquad -c_{+}\frac{2m_{H^+}^2}{v}hH^+H^- - c_{++}\frac{2m_{H^{++}}^2}{v}hH^{++}H^{--}\\
	&\qquad+\hat c_g\frac{\alpha_s}{12\pi v}hG_{\mu\nu}G^{\mu\nu} + \hat c_\gamma\frac{\alpha}{\pi v}h A_{\mu\nu}A^{\mu\nu},
\end{split}
\label{eq:hcouplings}
\end{equation}
where the $c_i$'s are coupling modifiers, $f$ stands for SM fermions, and $G_{\mu\nu}$ and $A_{\mu\nu}$ are the field strength tensors of the gluon and the photon, respectively.
Note that the couplings in the first two lines arise at tree level, while the ones in the last line arise at one-loop level.
In the SM we have $c^{SM}_W=c^{SM}_Z=c^{SM}_f=1$ and $\hat c_g^{SM}\simeq 0.97$ and $\hat c_\gamma^{SM}\simeq-0.81$. In the type-II seesaw model, these couplings  modifiers are
\begin{equation}
\begin{aligned}
	c_W &= \frac{\cos\alpha\, v_d + 2\sin\alpha\, v_t}{v},\\
	c_Z &= \frac{(\cos\alpha\, v_d + 4\sin\alpha\, v_t)v}{v^2+2v_t^2},
\end{aligned}\qquad
\begin{aligned}
	c_f &= \cos\alpha\frac{v}{v_d},\\
	\hat c_g &= c_f\hat c_g^{SM}.\phantom{\frac12}
\end{aligned}
\end{equation}
The expression for $\hat c_\gamma$ is more complicated. 
However, in the limit of small $v_t$ and small $\sin\alpha$, they reduce to
\begin{align}
	c_f &\simeq c_W\simeq c_Z \simeq 1,\quad  \hat c_g \simeq \hat c_g^{SM},\\
	\hat c_\gamma &\simeq \hat c_\gamma^{SM} +\delta \hat c_\gamma =\hat c_\gamma^{SM} + \frac{1}{24}c_{+}A_s\left(\frac{m_h^2}{4m_{H^+}^2}\right) + \frac{1}{6}c_{++}A_s\left(\frac{m_h^2}{4m_{H^{++}}^2}\right)\label{eq:deltacgamma},
\end{align}
where 
\begin{equation}
\begin{aligned}
	c_+ &\simeq \frac{(\lambda_1+\lambda_4/2)v^2}{2m_{H^+}^2},\qquad
	c_{++} \simeq \frac{\lambda_1v^2}{2m_{H^{++}}^2},
\end{aligned}
\label{eq:cgamma}
\end{equation}
and
\begin{equation}
\begin{aligned}
	A_s(\tau) &= \frac{3}{\tau^2}[F(\tau)-\tau],\\
	F(\tau) &= \left\{\begin{aligned}&\arcsin^2\sqrt{\tau},\\&-\frac{1}{4}\left[\log\frac{1+\sqrt{1-1/\tau}}{1-\sqrt{1-1/\tau}}-i\pi\right]^2,\end{aligned}\right.
	\quad\begin{aligned}&\tau\le1,\phantom{a^2}\\&\tau>1.\phantom{\left[\frac12\right]^2}\end{aligned}
\end{aligned}
\end{equation}
Thus we see that, in the small $v_t$ and small $\sin\alpha$ limit, the relevant Higgs data are the one concerning $h\to\gamma\gamma$ decay. Focusing on $h\to\gamma\gamma$, we define
\begin{equation}
	R_{\gamma\gamma} = \frac{\Gamma_{h\to\gamma\gamma}}{\Gamma_{h\to\gamma\gamma}^{SM}} \simeq 1 - 2.43\text{Re}(\delta\hat c_\gamma) + 1.50|\delta\hat c_\gamma|^2.
\label{eq:Rgammagamma}
\end{equation}
We will see in the next subsection that $R_{\gamma\gamma}$ plays an important role in determining the viable model parameter space.

\begin{table}
\begin{centering}
\begin{tabular}{| l | c | c | c |}
\hline
Search Channel & Tagged & Signal Strength  &Reference\\
\hline
$\gamma\gamma$ (ATLAS)& ggF & $0.81\pm0.19$ &  \cite{Aaboud:2018xdt}\\
  & VBF & $2.0\pm0.6$ &  \cite{Aaboud:2018xdt}\\
  & VH & $0.7\pm0.9$ &  \cite{Aaboud:2018xdt}\\
  & ttH & $0.5\pm0.6$ &  \cite{Aaboud:2018xdt}\\ \hline
$\gamma\gamma$ (CMS)& ggF & $1.10\pm0.19$ & \cite{Sirunyan:2018ouh}\\
  & VBF & $0.8\pm0.6$ &\cite{Sirunyan:2018ouh} \\
  & VH & $2.4\pm1.1$ & \cite{Sirunyan:2018ouh}\\
  & ttH & $2.2\pm0.9$ & \cite{Sirunyan:2018ouh}\\ \hline
  VV (ATLAS) & ttH & 1.5$\pm$0.6 & \cite{Aaboud:2017jvq} \\
ZZ (CMS) & ggF & 1.20$\pm$0.22 & \cite{Sirunyan:2017exp}\\
WW (CMS) & ggF & 1.35$\pm$0.20 &\cite{Sirunyan:2018koj}\\
 & ttH & 1.60$\pm$0.63 &\cite{Sirunyan:2018koj}\\ \hline
 $b\bar b$ (ATLAS) & ttH & $0.84\pm0.63$ & \cite{Aaboud:2017rss}\\
 $b\bar b$ (CMS) & VH & $1.2\pm0.4$ & \cite{Sirunyan:2017elk}\\
 & ttH & $0.72\pm0.45$ & \cite{Sirunyan:2018mvw} \\ \hline
 $\tau^+\tau^-$ (ATLAS) & ttH & $1.5\pm1.1$ &\cite{Aaboud:2017jvq} \\
 $\tau^+\tau^-$ (CMS) & ggF & $1.05\pm0.50$ &\cite{Sirunyan:2018koj} \\
 & VBF & $1.11\pm0.35$ & \cite{Sirunyan:2017khh} \\
 & ttH & $1.23\pm0.44$ & \cite{Sirunyan:2018shy}\\
\hline
\end{tabular}
\caption{The LHC Run II Higgs data employed in this work. The ``Tagged'' column displays the production mode of the Higgs boson: gluon fusion (ggF), vector boson fusion (VBF), associated production with a vector boson (VH) and associated production with a pair of $t\bar t$  (ttH). All uncertainties shown above are symmetrized.}
\label{tab:run2}
\end{centering}
\end{table} 

We use the $\chi^2$ fit to determine the compatibility of the type-II seesaw parameter space with the Higgs data. We construct the $\chi^2$ function based on both the LHC Run I and Run II Higgs data. For the Run I data, we use the official ATLAS and CMS measurements presented in table~8 of ref.~\cite{Khachatryan:2016vau}. For Run II, we include the data in the diphoton, diboson, $b\bar b$ and $\tau^+\tau^-$ search channels, see table~\ref{tab:run2}. 
In principle, the Higgs data for different production and decay channels are correlated. These correlations, however, are not readily extractable from the published measurements. Whenever available, we incorporate the correlations into our $\chi^2$ functions. This includes all the correlations for the Run I Higgs data as shown in figure 27 of ref.~\cite{Khachatryan:2016vau} and the Run II ATLAS diphoton measurements as shown in figure 40 of ref.~\cite{Aaboud:2018xdt}.

\subsection{Viable model parameter space}
\label{sec:viable}
In this subsection we investigate the viable parameter space consistent with perturbativity, vacuum stability, perturbative unitarity, the electroweak precision constraints and the Higgs data discussed above.  

We have seen in eqs.~\eqref{lambda}-\eqref{mH} that all quartic couplings are functions of triplet masses. Since those couplings are bounded by perturbativity, we would expect that those scalar masses will be constrained as well. Note, however, that some couplings also depend on the mixing angle $\alpha$ and/or the triplet vev $v_t$, see, e.g., eq.~\eqref{eq:lambda1}. The combination of these two parameters ($\alpha$ and $v_t$) proves to be central in constraining the parameter space of the model. For instance, when $\tan\alpha\ll v_t/v$, the second term of \eqref{eq:lambda1} becomes negligible. Perturbativity then implies that $m_{H^{++}}\leq\sqrt{2\pi}v$, which is numerically equal to 616 GeV. The bounds on other triplet masses can be easily found using eq.~\eqref{eq:mA}. 

In general, the situation is sometimes more complicated. An interplay among couplings often occurs, resulting in nontrivial  bounds. As an example, let us consider a case  where $\tan\alpha \simeq 3v_t/v$, giving $\lambda_1=(3m_h^2-6m_{H^+}^2+5m_{H^{++}}^2)v^{-2}$. Due to a negative sign in the second term,  $\lambda_1$ can be driven negative for $\Delta m>0$. Stability conditions, in particular condition (iii) of eq.~\eqref{eq:stability}, forbid $\lambda_1$ from being too negative. This, in turn, constrains the mass of the singly-charged scalar. In other words, it implies that $\Delta m$ decreases as $m_{H^{++}}$ increases. At some point, $\Delta m$ crosses zero and turns negative, so we have $\lambda_4<0$. As a result, condition (iv) of \eqref{eq:stability} becomes the one that gives a stronger constraint on the model  parameter space. One should note that both conditions (iii) and (iv) of \eqref{eq:stability} involve  $\lambda_2+\lambda_3$. Such a combination, crucial in determining the allowed parameter space of the model, is mostly constrained by eq.~\eqref{pert-uni}. 

Another interesting choice is $\tan\alpha=2v_t/v$. In this case, we have $\lambda_1+\lambda_4=\lambda$, which always satisfies condition (iv) of eq.~\eqref{eq:stability}. By using condition (iii), this yields an upper bound on $\lambda_4$, i.e., $|\lambda_4| < |\lambda + \sqrt{\lambda(\lambda_2+\lambda_3)}|$. Invoking eq.~\eqref{eq:lambda4}, we find that $m_{H^{++}}$ can be very large as long as $\Delta m$ is sufficiently small. This reflects the seesaw property of the model. One should notice that  $\tan\alpha=2v_t/v$ can also be obtained from eq.~\eqref{angle-even} with a very large triplet mass.

To identify the viable parameter space of the model, we perform the full parameter scan taking $v_t$, $\sin\alpha$, $m_{H^{++}}$,  $\Delta m$, $\lambda_2$ and $\lambda_3$ as free parameters.\footnote{The effect of $\lambda_2$ and $\lambda_3$ on the model phenomenology are minimal since they are suppressed by $v_t/v$.} In our scan, we take 
\begin{eqnarray*}
\centering
	v_t \in [10^{-9},5]\text{ GeV},\quad
	|\sin\alpha| \in &&[10^{-13},0.5],\quad
	m_{H^{++}} \in [60,1000]\text{ GeV},\\
	\Delta m \in [-50,50] \text{ GeV}, &&
	\lambda_2,\lambda_3 \in [-4\pi,4\pi].
\end{eqnarray*}  
We first generate 100,000 model parameter points compatible with the perturbativity, the vacuum stability, the perturbative unitarity and the electroweak constraints. From these points, we determine the minimum $\chi^2$ of the model to be 33.89. For comparison, the point corresponds to the SM has a $\chi^2$ value of 34.21. We also find that $|\sin\alpha|\lesssim 0.3$ at 95\% CL, reaffirming our identification of the $h$ with the 125-GeV resonance. As mentioned in the previous subsection, the parameter space compatible with the Higgs data is determined mostly by the $h\to\gamma\gamma$ decay. From our fit, we determine $0.81\le R_{\gamma\gamma}\le 1.27$ at 95\% CL. 

\begin{figure}[t]
       \centering
        \includegraphics[width= 0.475\textwidth]{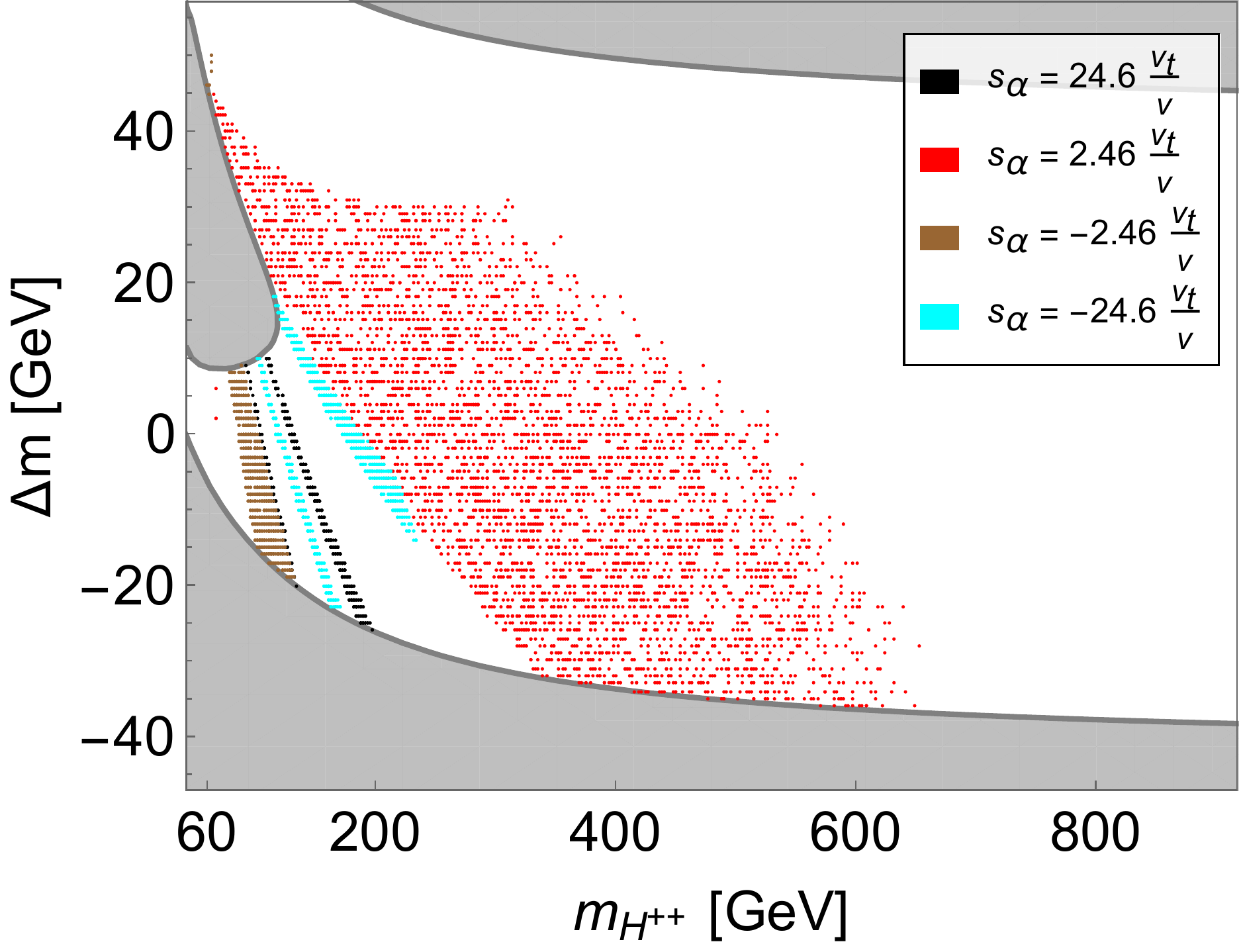}
        \hspace{.1cm}
        \includegraphics[width= 0.475\textwidth]{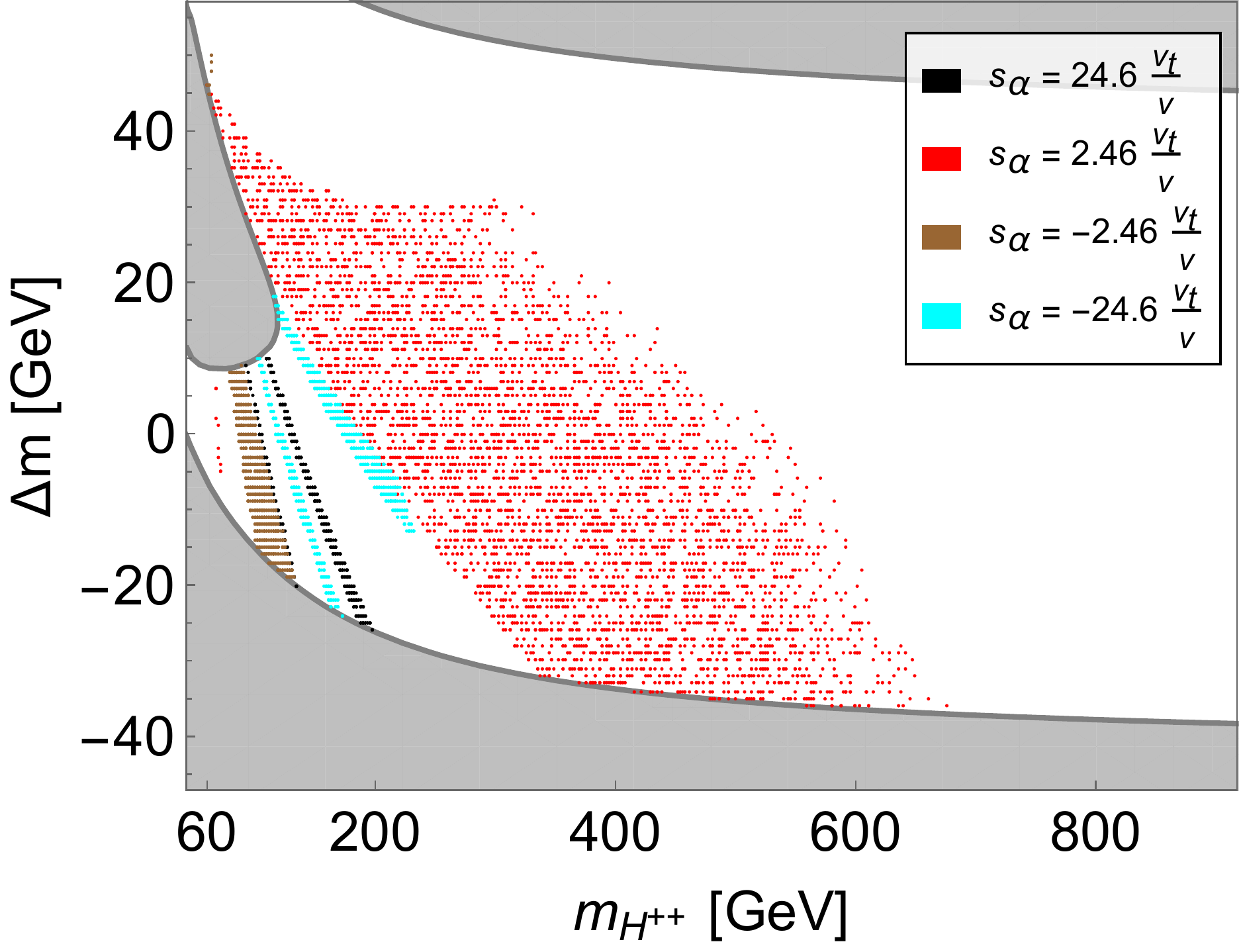}
        \caption{The region in the $m_{H^{++}}-\Delta m$ plane, consistent with the electroweak $S$ and $T$ parameters (open) and the 125-GeV Higgs data (scattered) at 95\% CL for the case $v_t=0.1$ GeV (left) and $v_t=10^{-9}$ GeV (right), with $v\sin\alpha/v_t = \pm2.46,\,\pm24.6$.}
         \label{fig:higgsdata}
\end{figure}

Next, we investigate the effect of each constraint on the allowed model parameter space. The $\rho$ parameter and our full parameter space scan imply that $v_t$ and $\sin\alpha$ are small. 
As a result, the scalar masses follow an approximate relations $m_H^2\simeq m_A^2\simeq  2m_{H^+}^2 - m_{H^{++}}^2$, see eqs.~\eqref{eq:mA} and~\eqref{mH}. Using this approximation, the $S$ and $T$ parameters in eq.~\eqref{eq:SandT} depend only on two parameters, $m_{H^{++}}$ and $\Delta m$. 
Thus we can identify the region in the $m_{H^{++}}-\Delta m$ plane inconsistent with the $S$ and $T$ parameters at 95\% CL. This is shown as a shaded region in figure~\ref{fig:higgsdata}. 

For the constraint imposed by the Higgs data, in particular $R_{\gamma\gamma}$, we note that in the small $v_t$ and $\sin\alpha$ regime, the couplings $c_+$ and $c_{++}$ depend approximately only on $m_{H^{++}}$, $\Delta m$ and the combination $v \sin\alpha/v_t$, see eqs.~\eqref{eq:lambda1} and~\eqref{eq:cgamma}. Thus we can derive a constraint on $m_{H^{++}}-\Delta m$ plane for a fixed value of $v\sin\alpha/v_t$. In figure~\ref{fig:higgsdata}, we show the parameter space consistent with the Higgs data, oblique parameters and theoretical constraints for four representative values of $v\sin\alpha/v_t = \pm2.46,\,\pm24.6$.

\begin{figure}[ht]
       \centering
        \includegraphics[width= 0.45\textwidth]{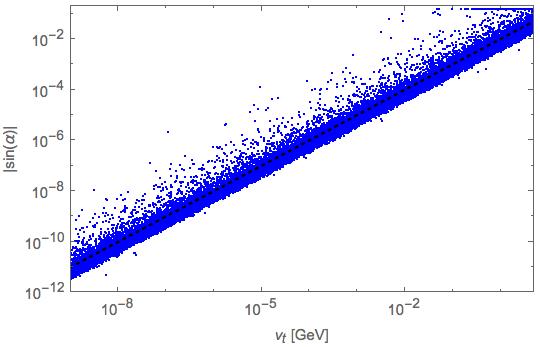}
        \hspace{.25cm}
        \includegraphics[width= 0.45\textwidth]{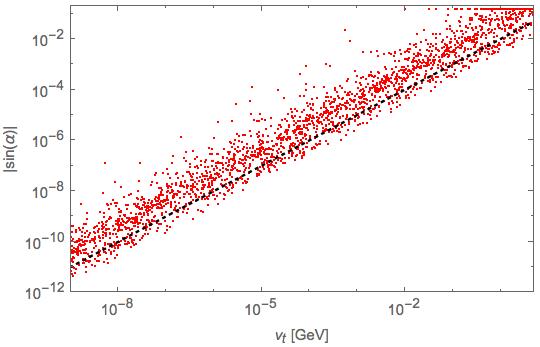}
        \caption{The viable parameter space in the $\sin\alpha$ vs $v_t$ plane. In the left figure $\sin\alpha$ is positive while in the right figure it is negative. In both cases, there is a clear correlation between the two parameters. On the dotted line  $|\sin\alpha|= 2.46 v_t/v$.}
         \label{fig:savt}
\end{figure}

%

One thing that is worth mentioning is that the allowed parameter space only depends on the {\it ratio} of $(v/v_t)\sin\alpha$. So, as long as the ratio is kept the same, such parameter space would not differ very much as shown in figure~\ref{fig:higgsdata}. There, one can see that the viable parameter space for the case $v_t=0.1$ and $10^{-9}$ GeV are almost identical. This is partly due to the fact that both $\sin\alpha$ and $v_t$ are correlated, see figure~\ref{fig:savt}.


\subsection{Lepton flavor violation constraints}
\label{sec:LFV}
The Yukawa interactions presented in eq.~(\ref{yuk}) will lead naturally to flavor-changing leptonic decays such as  $\ell_a^- \to \ell_b^+\ell_c^-\ell_d^-$ and $\ell_a^- \to \ell_b^-\gamma$. The former arises at tree level and is mediated by the doubly-charged Higgs $H^{++}$. Its rate is found to be
\begin{eqnarray}
\Gamma(\ell_a^- \to \ell_b^+\ell_c^-\ell_d^-) = \frac{1}{2(1+\delta_{cd})}\frac{m_{\ell_a}^5}{192\pi^3}\left|\frac{f_{ab}f_{cd}}{m_{H^{++}}^2}\right|^2,
\label{mu-3e}
\end{eqnarray} 
where the Kronecker delta $\delta_{cd}$ accounts for possible two identical final states. 

The $\ell^-_a \to \ell_b^-\gamma$ decay occurs through one-loop penguin diagrams, mediated either by the $H^+$ or the $H^{++}$. In a process mediated by the $H^+$, the photon is emitted only from the $H^+$, 
whereas in a process mediated by the doubly-charged Higgs, the photon can be emitted from either the $H^{++}$ or the charged fermion propagating inside the loop. Since  both scalars in this model couple to leptons with the same chirality, their contributions coherently interfere, yielding 
\begin{eqnarray}
\Gamma(\ell^-_a \to \ell^-_b\gamma) = \frac{m_{\ell_a}^5\alpha_{em}}{(192\pi^2)^2} |f^\dagger f|_{ab}^2 \left(\frac{1}{m_{H^+}^2} + \frac{8}{m_{H^{++}}^2} \right)^2.
\label{mu-e-gamma}
\end{eqnarray}
In deriving this equation, we have neglected all internal lepton masses because they are much smaller than $m_{H^+}$ or $m_{H^{++}}$. $\alpha_{em}$ is the electromagnetic fine-structure constant. In our calculation we take $\alpha^{-1}_{em}(m_Z)=128$. Note the factor of 8 in the $H^{++}$ term. A portion of it comes from the normalization factor of $\Delta^+$ (which will become $H^+$), see eq.~(\ref{sc}), leading to relative enhancement for $H^{++}$ term by a factor of 2. 
The rest is due to the fact that the loop function of diagram with photon being emitted from the internal fermion $\ell^c$ is two times larger than the one with photon being emitted from the $H^{++}$. But the electric charge of the fermion is two times smaller than that of $H^{++}$, resulting in the additional factor of 4 at the amplitude level.

\begin{table}
\begin{center}
\begin{tabular}{lcc}
\hline\hline
Process & Branching ratio bound & Constraint \\
\hline 
$\mu^- \to e^+e^-e^-$ & $1.0\times 10^{-12}$ & $ m_{H^{++}}> |(M_\nu)_{\mu e}(M_\nu)_{ee}|^{1/2}/v_t \times {145~\rm TeV}$\\
$\tau^- \to e^+e^-e^-$ & $2.7\times 10^{-8}$ & $ m_{H^{++}}> |(M_\nu)_{\tau e}(M_\nu)_{ee}|^{1/2}/v_t \times {7.4~\rm TeV}$\\
$\tau^- \to e^+e^-\mu^-$ & $1.8\times 10^{-8}$ & $ m_{H^{++}}> |(M_\nu)_{\tau e}(M_\nu)_{e\mu}|^{1/2}/v_t \times {9.8~\rm TeV}$\\
$\tau^- \to e^+\mu^-\mu^-$ & $1.7\times10^{-8}$ & $ m_{H^{++}}> |(M_\nu)_{\tau e}(M_\nu)_{\mu\mu}|^{1/2}/v_t \times {8.3~\rm TeV}$ \\
$\tau^- \to \mu^+e^-e^-$ & $1.5\times 10^{-8}$ &$ m_{H^{++}}> |(M_\nu)_{\tau\mu}(M_\nu)_{ee}|^{1/2}/v_t \times {8.6~\rm TeV}$  \\
$\tau^- \to \mu^+\mu^-e^-$ & $2.7\times 10^{-8}$ & $ m_{H^{++}}> |(M_\nu)_{\tau\mu}(M_\nu)_{\mu e}|^{1/2}/v_t \times {8.8~\rm TeV}$ \\
$\tau^- \to \mu^+\mu^-\mu^-$ & $2.1\times 10^{-8}$ & $ m_{H^{++}}> |(M_\nu)_{\tau\mu}(M_\nu)_{\mu\mu}|^{1/2}/v_t \times {7.9~\rm TeV}$\\
$\mu\to e\gamma$ & $4.2\times 10^{-13}$ & $m_{H^{++}}> \sqrt{(8+r)|M_\nu^\dagger M_\nu|_{\mu e}}/v_t \times 15.3~\rm TeV$\\
$\tau\to e\gamma$ & $3.3\times 10^{-8}$ & $m_{H^{++}}> \sqrt{(8+r)|M_\nu^\dagger M_\nu|_{\tau e}}/v_t \times  0.6~\rm TeV$\\
$\tau\to \mu\gamma$ & $4.4\times 10^{-8}$ & $m_{H^{++}}> \sqrt{(8+r)|M_\nu^\dagger M_\nu|_{\tau\mu}}/v_t \times 0.56~\rm TeV$\\
\hline\hline

\end{tabular}
\caption{List of various LFV processes constraining the model. Here we use $M_\nu=\sqrt{2}fv_t$ to express Yukawa couplings $f_{ab}$ in terms of neutrino mass matrix elements. We also define $r\equiv m_{H^{++}}^2/m_{H^+}^2$. All bounds are taken from ref.~\cite{Tanabashi:2018oca}.}
\label{lfv}
\end{center}
\end{table}

Current bounds for these flavor-violating processes are presented in table~\ref{lfv}. 
The two most constraining processes are $\mu\to3e$ and $\mu\to e\gamma$, from which we can derive the lower bound on the mass of the triplet. It is clear that the bound will depend on the value of $v_t$ and the type of neutrino mass hierarchy. 
For illustration purpose, let us consider a case where $v_t=10^{-9}$ GeV with the normal neutrino mass hierarchy and $m_1=0$. 
We use the neutrino mass matrix given in eq.~(\ref{numass-nh}). For simplicity, we further assume degenerated triplet masses, and thus $r=m_{H^{++}}^2/m_{H^+}^2=1$. 
(Allowing non-degenerated masses will merely induce a 10\% correction, thanks to the $T$ parameter limiting the mass splitting of the triplet.) We infer from table~\ref{lfv} that $\mu\to 3e$ and $\mu\to e\gamma$ set $m_{H^{++}}> 510$ GeV and $m_{H^{++}}>750$  GeV, respectively. Comparing this result with the allowed region shown in figure \ref{fig:higgsdata}, we conclude that the benchmark choices in figure \ref{fig:higgsdata} are incompatible with LFV for $v_t=10^{-9}$ GeV. But it should be kept in mind that other benchmark choice, e.g., $\sin\alpha=2v_t/v$, corresponding to the seesaw limit, is still compatible with LFV constraints because the triplet masses are allowed to be large. Bounds for other $v_t$ are also determined in the same way. One should note that  for larger $v_t$, the obtained bounds become weaker by a factor of $v_t/(10^{-9}~\rm GeV)$, see the third column of table~\ref{lfv}.

We see, for this case, that the  $\mu\to e\gamma$ gives a more stringent constraint compared to $\mu\to3e$ decay despite being originated radiatively.  This is because  the $\mu\to e\gamma$ rate contains terms that are elements of (2,3) sector of $M_\nu$, which are dominant entries in NH. The $\mu\to3e$ decay rate, on the other hand, involves a product of much smaller elements, i.e., $(M_\nu)_{ee}$ and $(M_\nu)_{\mu e}$, resulting in a weaker limit.

The situation will be different if we consider a NH case with $m_1=0.1$ eV. (Note that the three neutrino masses are almost equal here.) Now the main diagonal entries of $M_\nu$ are much larger than the off-diagonal parts. So the couplings involved in both processes are roughly of the same order, but the $\mu\to e\gamma$ is suppressed by a loop factor. This is the reason why the tree-level decay $\mu\to 3e$ sets a stronger limit, $m_{H^{++}}>5.5$ TeV, compared to $m_{H^{++}}>950$ GeV from the  $\mu\to e\gamma$ process. Similar argument also applies for the IH case. The bounds here are much stronger, eliminating the few-hundred-GeV parameter space of the model with $\sin\alpha$ assumed to be $2.46v_t/v$. We find $m_{H^{++}} > 2.38$ TeV for $m_3=0$ and $m_{H^{++}}>7.3$ TeV for   $m_3=0.1$ eV, respectively. 

It is worth noting that the entire bounds discussed in this section are derived based on the neutrino mass matrix given in eqs.~(\ref{numass-nh}) and (\ref{numass-ih}), where all neutrino oscillation parameters are taken to be their best-fit values, while Majorana phases are set to zero. In principle, varying each parameter within its allowed range will lead to different bounds, which could be weaker in some cases. For example, allowing all parameters to vary within their $3\sigma$ range will push the doubly-charged mass to about 650 GeV for NH case. 
Upon closer inspection, we find the bound is mainly driven by $\theta_{23}$. The above bound is achieved by lowering $\theta_{23}$ to its $2\sigma$ lower bound. 

As for the IH, the lower bound can even be as low as 510 GeV, thereby allowing quite a significant parameter space from the Higgs data and vacuum stability. However, in contrast to the NH case, it seems that more parameters are involved. This is due to the fact that in the IH scenario there are two dominant neutrino masses, i.e., $m_1$ and $m_2$, which are narrowly split. Thus, a slight change on neutrino mass parameters may lead to a totally different result. 

\section{Collider phenomenology}
\label{sec:ColliderBounds}
In this section, we turn our attention to constraints placed on the type-II seesaw model by collider experiments. In particular, the extra scalar bosons can be produced and searched for at the LHC. As discussed in the previous sections, the relevant parameter space of the model  consists of $v_t\ll v$ and small $\sin\alpha$. 
This leads to an approximate relation on the scalar masses as shown in eqs.~\eqref{eq:mA} and~\eqref{mH}. Thus, for simplicity, we will assume $m_{H}^2 = m_{A}^2 = 2 m_{H^{+}}^2 - m_{H^{++}}^2$ for the rest of this section. 

There are several official LHC searches for the charged and the neutral scalar bosons. Most of the searches, however, are in the context of two Higgs doublet model. For the singly-charged Higgs searches, both the CMS and the ATLAS are interested in the $g g \rightarrow \bar t b H^+$ and $g \bar b \rightarrow \bar t H^+$ production channels~\cite{Khachatryan:2015qxa, CMS:2018ect, CMS:2016szv, Aaboud:2018gjj, Aaboud:2018cwk}. These production channels depend on the $t b H^+$ coupling, which is suppressed by $\tan^2\beta = 2 v_t^2 / v_d^2 \lesssim \mathcal O(10^{-4})$. Hence the bounds imposed from these LHC searches do not apply in our case. Additionally the $H^+$ is also searched in the vector boson fusion production channel~\cite{Aaboud:2018ohp, Sirunyan:2017sbn}. Moreover, for $m_{H^+} < m_t - m_b$, the $H^+$ can be searched for in the decay of $t \rightarrow b H^+$~\cite{Sirunyan:2018dvm, Khachatryan:2015uua, CMS:2018ect, CMS:2016szv, Aaboud:2018gjj, Aad:2013hla}. As in the previous case, these searches also suffer from $\tan^2\beta$ suppression, rendering them irrelevant for constraining the type-II seesaw model. 

In the case of neutral scalars $\phi=H$ and $A$, the official ATLAS and CMS searches concentrate on the gluon fusion production~\cite{Aaboud:2017sjh, Sirunyan:2018zut, Aaboud:2018bun}, associated with $b$-jet productions~\cite{Aaboud:2017sjh, Sirunyan:2018zut, Sirunyan:2018taj} and associated production with a vector boson~\cite{Aaboud:2018bun}. These production channels are suppressed by $\sin^2\alpha$ and/or $\sin^2 \beta'$. Hence, the bounds derived by the official ATLAS and CMS searches for neutral scalars are not applicable to the type-II seesaw model.

The doubly-charged Higgs has also been searched for by the ATLAS and CMS collaborations. Unlike the case of the neutral and the singly-charged Higgs, the official searches focus on the $H^{++}$ produced via the Drell-Yan process, which is not suppressed by $\sin\alpha$ or $\tan\beta$. ATLAS have been searching for $H^{++}\to W^+W^+$~\cite{Aaboud:2018qcu} and $H^{++}\to\ell^+\ell^+$~\cite{Aaboud:2017qph} from the pair production of $H^{++}H^{--}$ using 36.1 fb$^{-1}$ of data at 13 TeV. For the $W^+W^+$ channel, ATLAS have established a lower bound $m_{H^{++}}\ge 220$ GeV. For the $\ell^+\ell^+$ channel, the bound depends on the assumed branching ratio of $H^{++}\to\ell^+\ell^+$. It ranges from $m_{H^{++}}\ge 450$ GeV for $Br(H^{++}\to \ell^+\ell^+)=10\%$ to $m_{H^{++}}\ge 770$ GeV for $Br(H^{++}\to \ell^+\ell^+)=100\%$. We will refer to these bounds as the official ATLAS same-sign diboson and same-sign dilepton bounds, respectively. Meanwhile, CMS have searched for $H^{++}\to\ell^+\ell^+$ in the $H^{++}H^{--}$ and $H^{++}H^{-}$ production channels using 12.9 fb$^{-1}$ of data at 13 TeV~\cite{CMS:2017pet}. As in the case with ATLAS, the official bounds quoted by CMS depend strongly on the assumed branching ratios of $H^{++}$ into same-sign dilepton. We will refer to these bounds as the official CMS same-sign dilepton bounds.

\begin{figure}[t!]
       \centering
        \includegraphics[width= 0.7\textwidth]{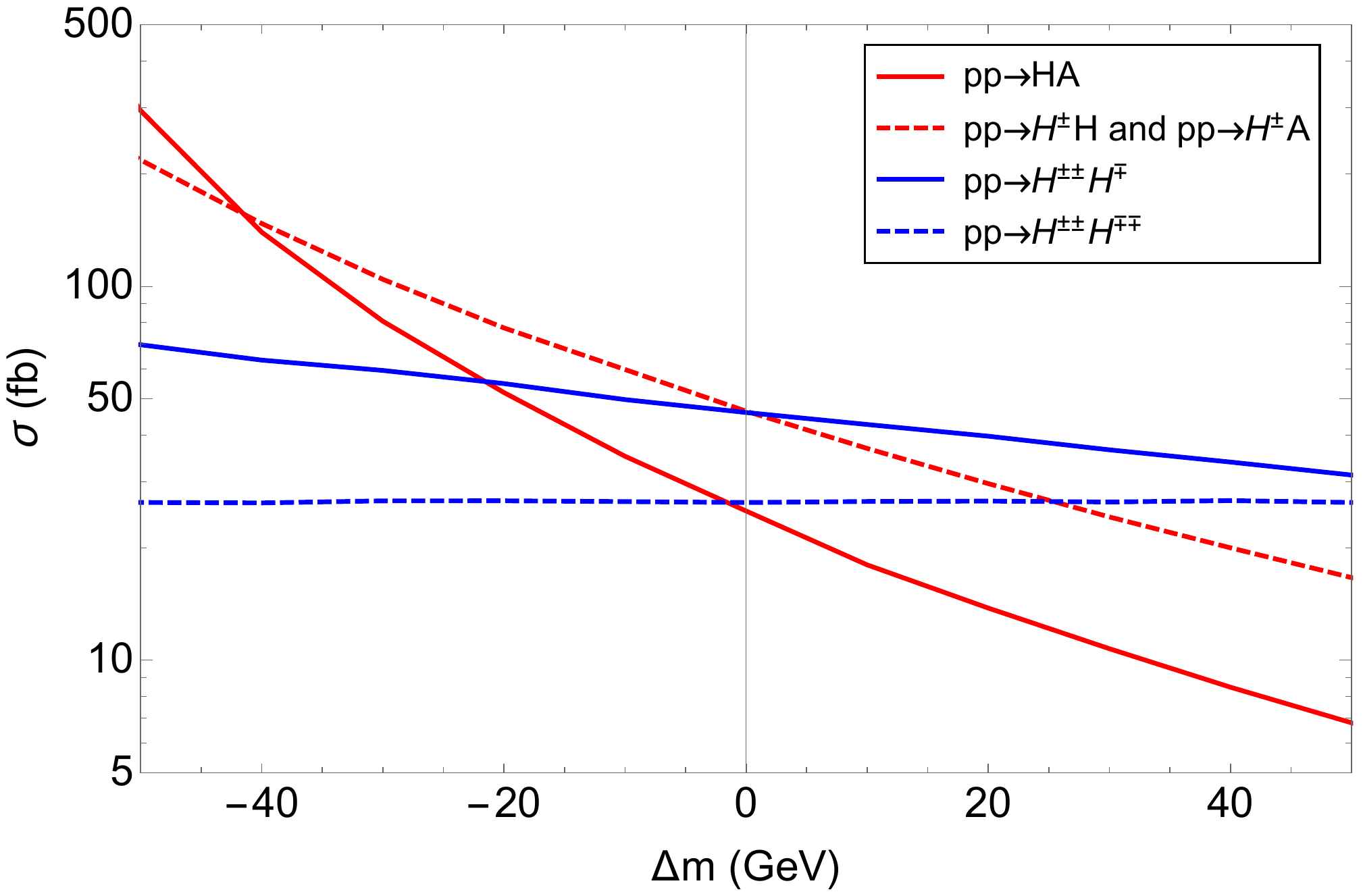}
        \caption{The production cross sections for each production channel for $m_{H^{++}} = 250$ GeV as a function of the mass difference $\Delta m$. In this figure we assume $m_{H}^2 = m_{A}^2 = 2 m_{H^{+}}^2 - m_{H^{++}}^2$.         }
         \label{fig:scalarxsection}
\end{figure}

\begin{figure}
       \centering
                \subfloat[$H$]{\includegraphics[width= 0.47\textwidth]{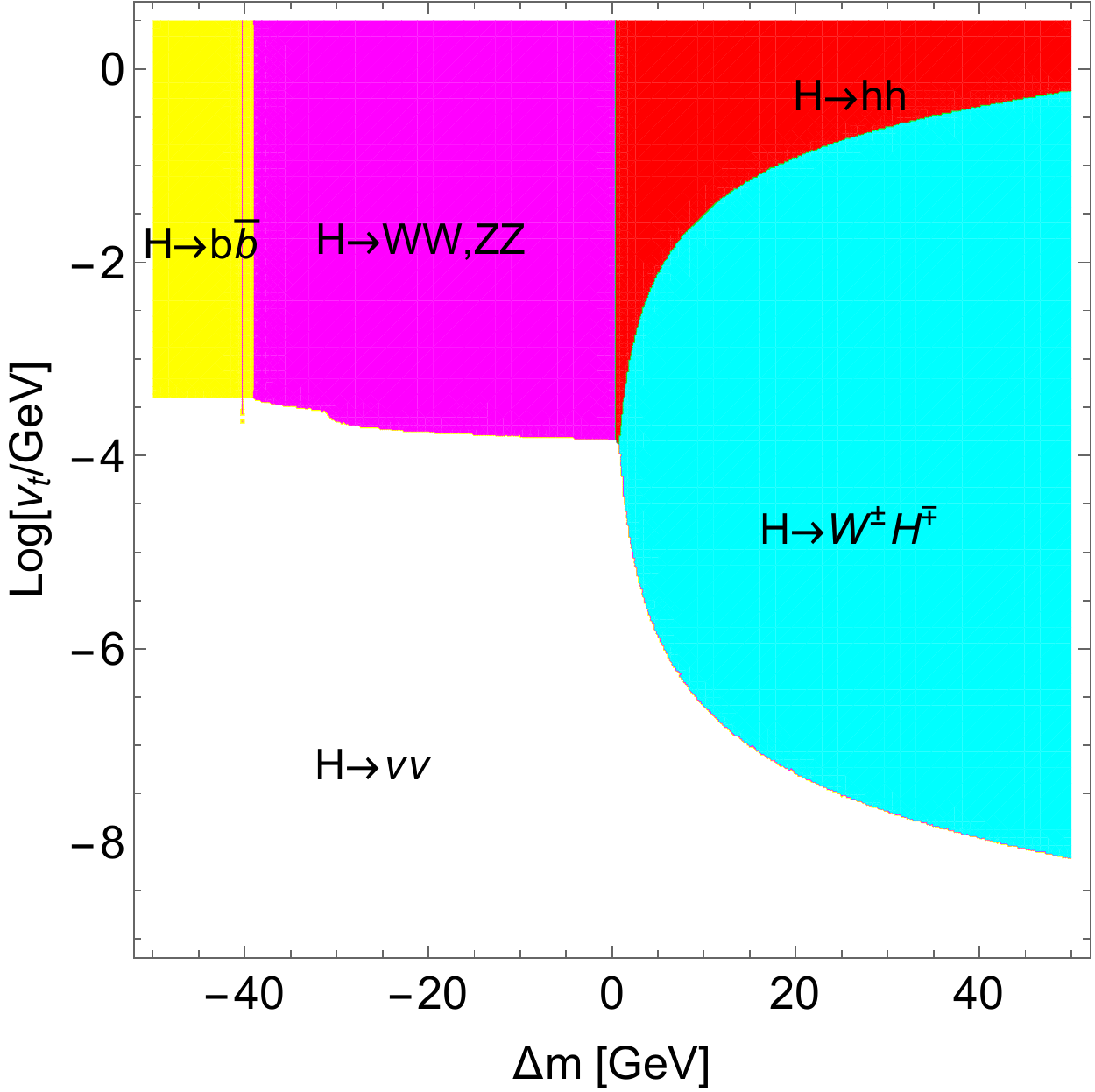}\label{fig:hvtdm200}}  \,\,\,
        \subfloat[$A$]{\includegraphics[width= 0.47\textwidth]{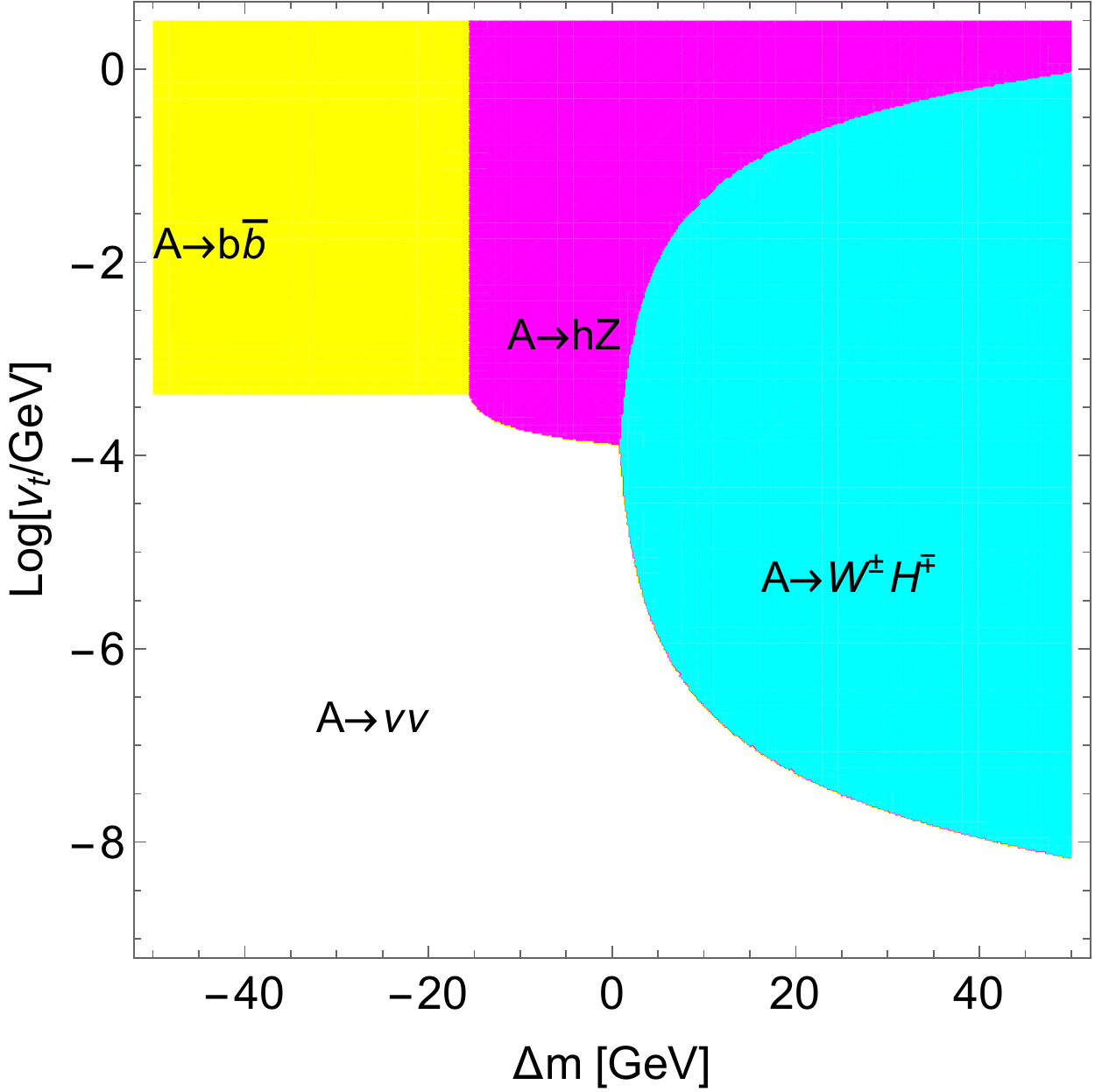}\label{fig:avtdm200}}  \\
                        \subfloat[$H^+$]{\includegraphics[width= 0.47\textwidth]{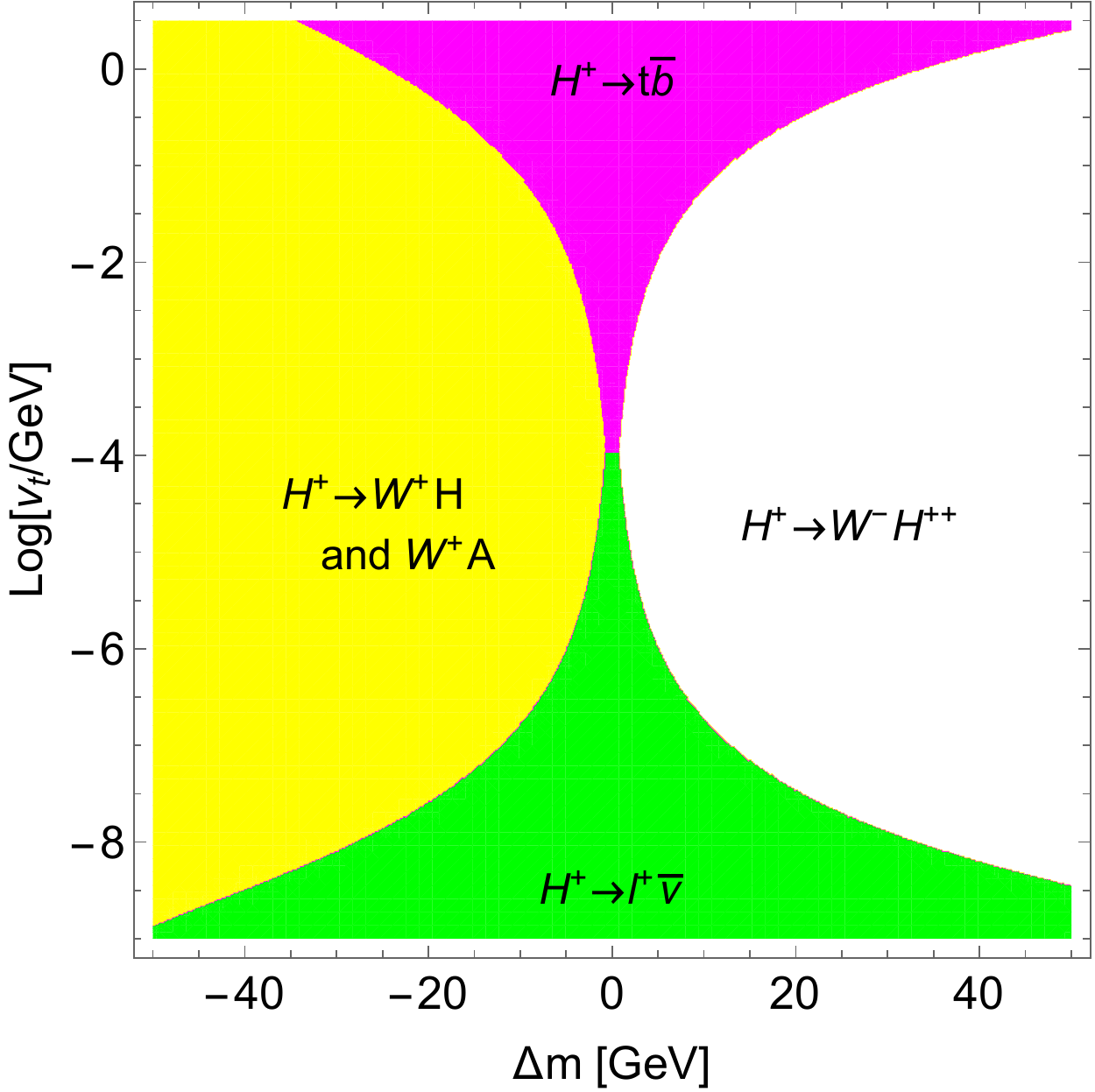}\label{fig:hpvtdm200}}  \,\,\,
        \subfloat[$H^{++}$]{\includegraphics[width= 0.47\textwidth]{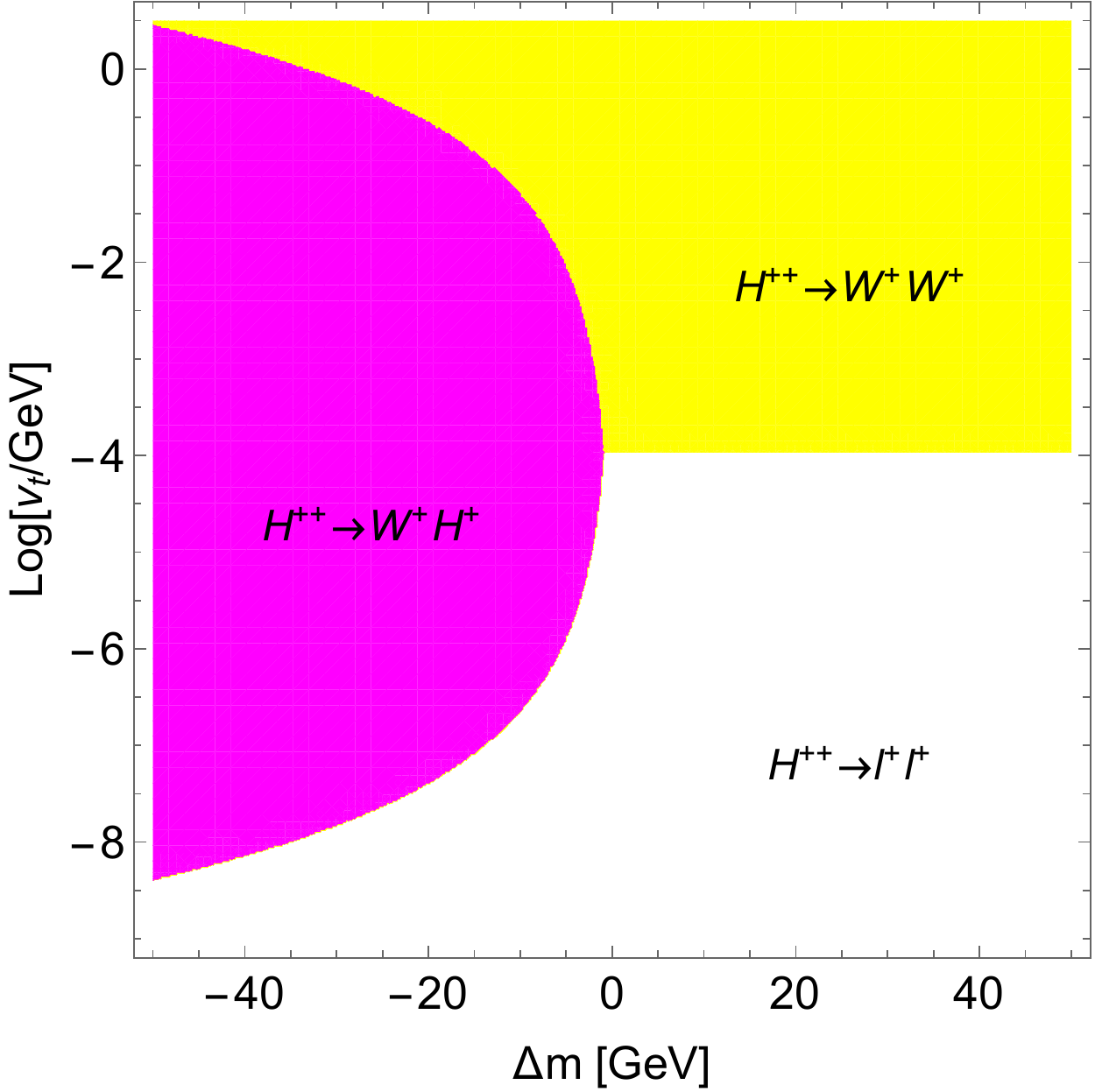}\label{fig:hppvtdm200}} 

        \caption{The channels with largest partial decay width for the scalars in the plane of $\Delta m$--$v_t$ for $m_{H^{++}} = 250$ GeV. Here we take $s_\alpha= 2.46 v_t/v$. We assume $m_{H}^2 = m_{A}^2 = 2 m_{H^{+}}^2 - m_{H^{++}}^2$. In the case of heavier $H$, the region of $H \rightarrow hh$ gets bigger and replaces $H \rightarrow WW,ZZ$ and $H \rightarrow b \bar b$ since more phase space opens. A similar scenario also happens as $A$ gets heavier, in which the $A \rightarrow h Z$ region replaces the $A \rightarrow b \bar b$. In the case of heavier $H^+$, the regions $H^+ \rightarrow W^+ Z$ and $H^+ \rightarrow t \bar b$ get replaced by $H^+ \rightarrow W^+ h$, subject to phase space allowance. The largest channel for $H^{++}$ does not significantly change as the mass of $H^{++}$ gets heavier. }
         \label{fig:scanvtdm200}
\end{figure} 

It is worth noting that the official ATLAS and CMS searches for $H^{++}$ do not cover all the relevant parameter space of the current model. The results quoted above only represent the extreme corners of the model parameter space, i.e., high and low $v_t$ regimes. Besides, they do not take into account all possible production channels and decay modes, which may be relevant in constraining the full parameter space of the model. As already noted, the relevant production mechanism for the scalars is $\phi\phi'$ (with $\phi,\phi'\in\{H,A,H^+,H^{++}\}$) via the Drell-Yan process. Since the splitting among scalar masses occurs according to eqs.~\eqref{eq:mA} and \eqref{mH}, we can write one mass in terms of the others. Thus, we can express the production cross-sections as functions of two model parameters only: $\Delta m$ and $m_{H^{++}}$.  

In figure~\ref{fig:scalarxsection}, we present four production cross-sections at the LHC 13 TeV as functions of scalar mass splitting $\Delta m$, with the doubly-charged mass $m_{H^{++}}$ fixed at 250 GeV.
Note that the cross sections in this case are $\mathcal O(100)$ fb, suggesting a significant number of them being produced. From that figure, we can also see that the production cross-sections for $H^\pm H$ and $H^\pm A$, not considered in both ATLAS and CMS analyses, are sizable, and even larger than $H^{\pm\pm}H^{\mp}$ and $H^{\pm\pm}H^{\mp\mp}$, particularly in $\Delta m<0$ region. This is because the neutral scalars are lighter than their charged counterparts. This indicates that the $H^\pm H$ and $H^\pm A$ production channels need to be included in the analysis. As we will see later, the inclusion of these two channels, along with the possible decay modes, will play an important role in deriving the LHC bounds for $\Delta m<0$. On the other hand, for $\Delta m > 0$, the pair productions of $H^{\pm\pm} H^{\mp\mp}$ and $H^{\pm\pm} H^\mp$ are the two most important channels for this purpose.

In addition to the production mechanism, the LHC signatures also depend on the decay channels of each scalar boson being produced. Within this model, the scalar decay modes can be affected by four parameters: $\Delta m$, $m_{H^{++}}$, $v_t$ and $\sin\alpha$. 
Plots in figure~\ref{fig:scanvtdm200}
describe the main decay channels for each scalar as functions of $v_t$ and $\Delta m$ with $m_{H^{++}}$ fixed at 250 GeV. For this illustrative purpose, the mixing angle $\alpha$ is set to $s_\alpha=2.46v_t/v$ because, as we have learned from figure~\ref{fig:savt}, both $s_\alpha$ and $v_t$ are correlated. The proportionality factor of 2.46 comes from the fact that this particular choice gives quite large allowed parameter space (see figure~\ref{fig:higgsdata}), although practically this makes the mixing angle nearly vanishing. We also assume that the triplet Yukawa couplings, relevant for small $v_t$ cases ($v_t\lesssim10^{-7}$ GeV), are chosen such that they are consistent with neutrino mass matrix presented in eq.~\eqref{numass-nh}, i.e., NH with  $m_1=0$.

In what follows, we will investigate how collider searches for the scalars can further probe the parameter space of the type-II seesaw model. We will utilize all the relevant production channels and decay modes discussed above. And since we will consider more general aspect of the model, we cannot completely be reliant on those two official results. Because of that, due to the great similarities in the production and decay modes, we will also implement the CMS multilepton analysis~\cite{Sirunyan:2017lae}, intended to constrain supersymmetric particles.  To this end, we first use Madgraph5~\cite{Alwall:2014hca} to generate these scalars. We then pass them through Pythia8~\cite{Sjostrand:2014zea} for showering and hadronization. Finally, we use the Checkmate2~\cite{Dercks:2016npn} public code to obtain the relevant LHC constraints. 

\begin{figure}
       \centering
                \includegraphics[width= 0.8\textwidth]{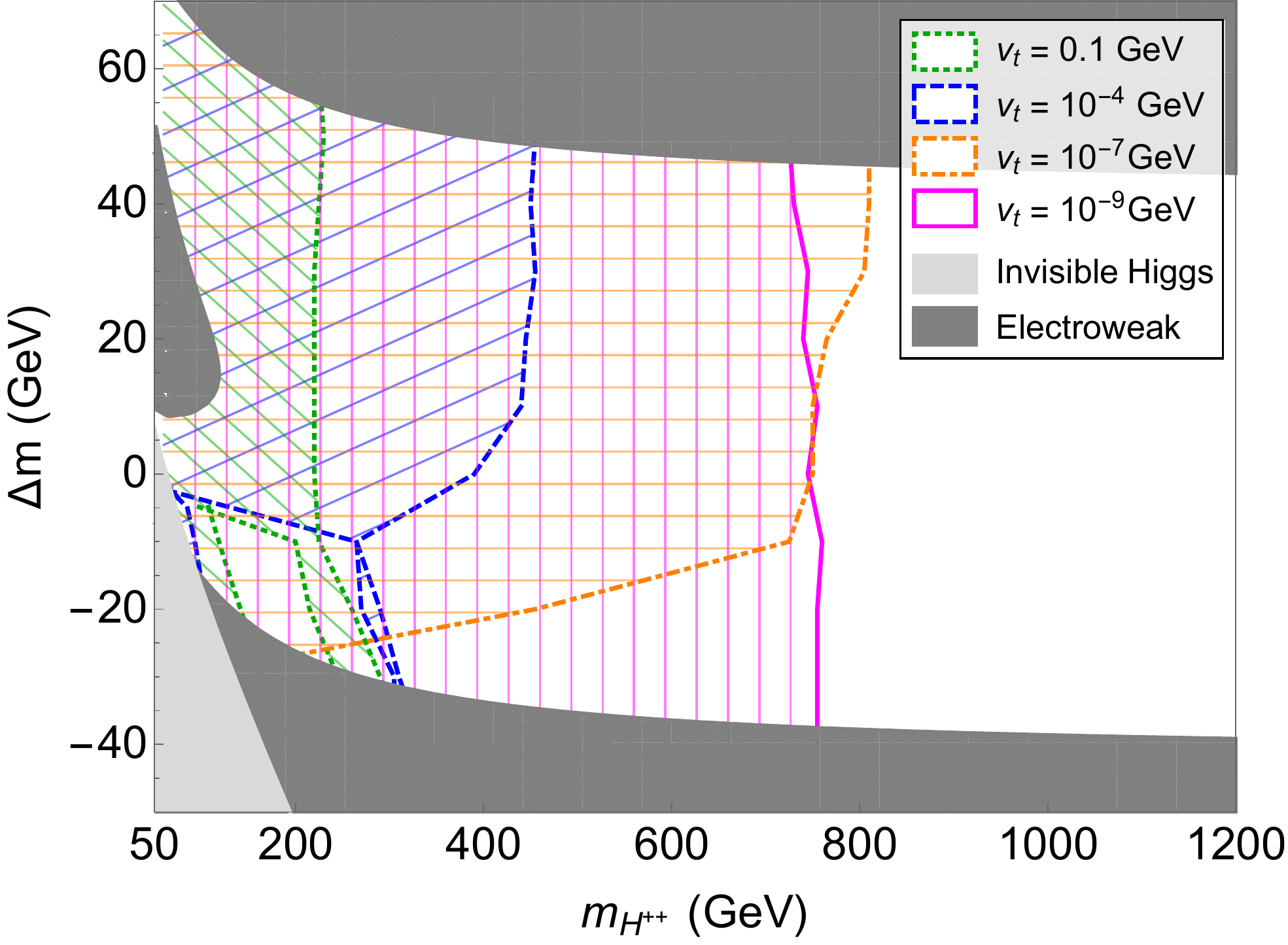}
        \caption{The summary of the bounds from collider, invisible Higgs decay and the electroweak precision test. The excluded regions from collider searches are represented by hashed green, blue, orange and magenta regions for $v_t = 0.1$ GeV, $10^{-4}$ GeV, $10^{-7}$ GeV and $10^{-9}$ GeV respectively. The invisible Higgs and electroweak bounds are independent of $v_t$ and represented by light and dark grey shaded regions respectively.}
         \label{fig:allbounds}
\end{figure}

In the next two subsections, we will discuss how to get the collider bounds, and hence the constraint for the parameter space. We will present bounds for $v_t=0.1,~10^{-4},~10^{-7},~10^{-9}$~GeV that can encompass wide range of the model parameter space.  Due to different nature in production and decay processes, we split our discussion into $\Delta m>0$ and $\Delta m<0$. The bounds are summarized in figure~\ref{fig:allbounds}.

\subsection{$\Delta m > 0$}

\begin{figure}
       \centering
        \includegraphics[width= 0.7\textwidth]{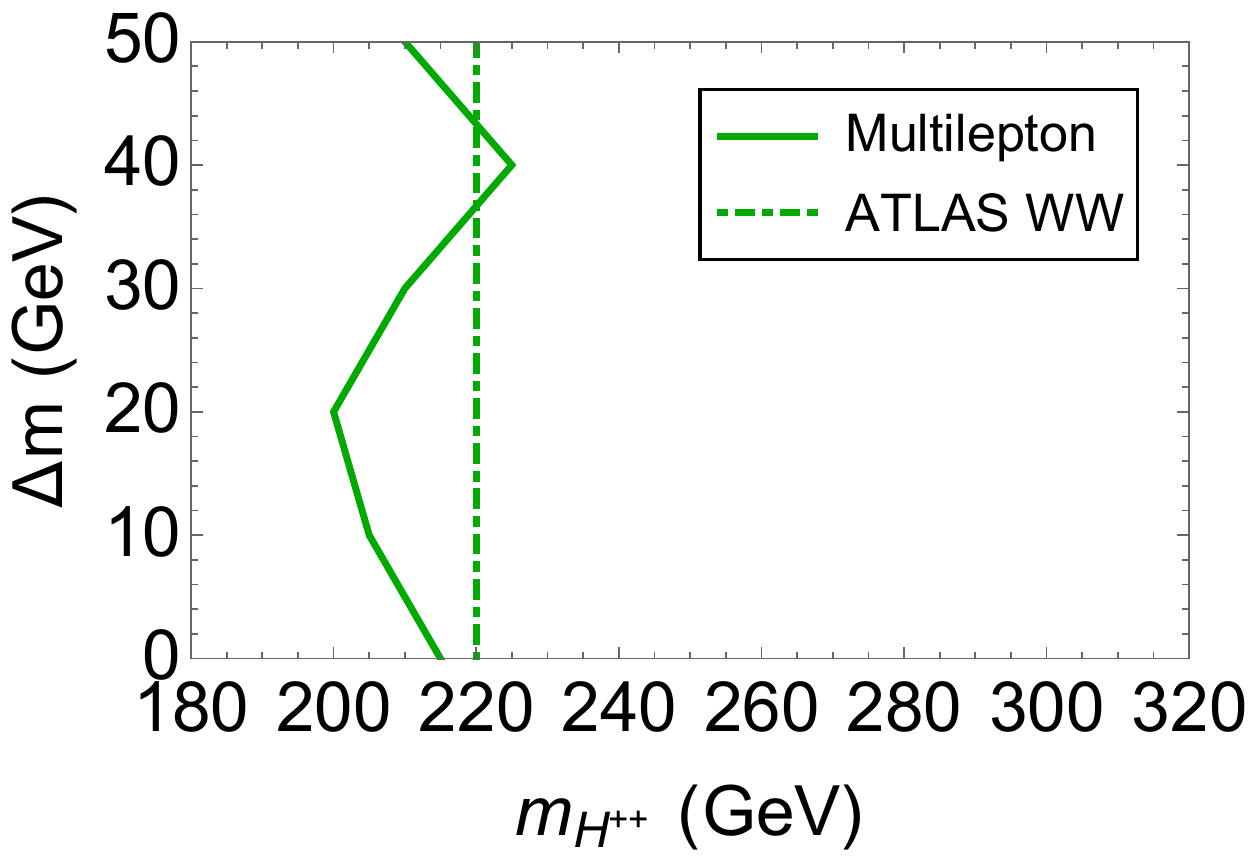}
        \caption{The comparison between the ATLAS same sign diboson bounds~\cite{Aaboud:2018qcu} and the bounds from recasting the CMS multilepton analysis~\cite{Sirunyan:2017lae} for $v_t = 0.1$ GeV. The region on the left of the respective line is excluded.}
         \label{fig:wwvsmultilepton}
\end{figure}

For  $\Delta m > 0$, the charged Higgses are most likely to get produced. Thus, we focus our attention on the LHC signatures arising from $H^{++}$ and $H^+$ decays.

First, let us consider the case of large $v_t$ in which the $H^{++}$ decays exclusively into $W^+W^+$. In this case, the official ATLAS same-sign diboson bound, where only the pair production of $H^{++}H^{--}$ is considered, is $m_{H^{++}}\ge 220$ GeV. We will investigate the effect of adding the $H^{++}H^-$ production channel, which is the most dominant production mechanism for $\Delta m>0$, on the extraction of $m_{H^{++}}$ bound. To do this, we note that, for small $\Delta m$, the $H^+$ decays predominantly into $W^+Z$, see figure~\ref{fig:hpvtdm200}. 
Thus, the final state relevant for the $H^{++}H^-$ production channel is $W^+ W^+ W^- Z$. This suggests that the $H^{++}H^-$ channel can be constrained by multilepton search. We recast the CMS multilepton search at 13~TeV with 35.9~fb$^{-1}$ of data~\cite{Sirunyan:2017lae} using Checkmate2~\cite{Dercks:2016npn}.\footnote{We also consider other channels available in Checkmate2. We find that only multilepton channel is relevant for constraining the model.} We find that the signal region G05, defined in table~\ref{tab:cuts}, which involves four or more leptons and a large missing energy, provides the strongest bounds. The comparison between the recasted CMS multilepton bounds and the official ATLAS same-sign diboson bounds, for the case of $v_t = 0.1$ GeV, is shown in figure~\ref{fig:wwvsmultilepton}. One can see that for this value of $v_t$, the multilepton bounds are comparable with the same-sign diboson bound.


Next, we consider the case of tiny $v_t$, in which $H^{++}$ decays predominantly into a pair of same-sign dilepton. In the official ATLAS same-sign dilepton search~\cite{Aaboud:2017qph}, the bounds are presented for the following decay channels: $e^+e^+$, $\mu^+\mu^+$ and $e^+\mu^+$. Reinterpreting these bounds in the context of eq.~\eqref{numass-nh} with $m_1=0$, we find that the strongest bounds come from the $\mu^+\mu^+$ channel, giving $m_{H^{++}}\ge 667$ GeV for both $v_t=10^{-7}$ GeV and $v_t=10^{-9}$~GeV. Since ATLAS did consider possible decay into states other than light leptons, as $v_t$ gets larger, the leptonic modes naturally get smaller, leading to relatively weaker bound. As an example, for $v_t=10^{-4}$~GeV, the bound becomes $m_{H^{++}}\gtrsim 450$ GeV. Note that we do not consider the official CMS same-sign dilepton search here because it was performed on a smaller LHC data set and are more difficult to recast. We will discuss, however, some CMS  benchmark points later in the appendix.


\begin{table}[t!]
\centering
\begin{tabular}{||c | c | c||} 
 \hline
 SR  & A44 & G05 \\ [0.5ex] 
 \hline\hline
 $p_{T, \text{ leading electron (muon)}}$& 25(20) GeV & 25(20) GeV \\
  $p_{T, \text{ subleading electrons (muons)}}$& 15(10) GeV & 15(10) GeV \\
 $n_\ell $ & 3 & $\geq 4$ \\ 
 $p_{T, \tau_h}$ & 20 GeV & 20 GeV \\
  $n_{\tau_h} $ & 0 & 0 \\ 
 $p_{T,b\text{-jet}}$ & 25 GeV & 25 GeV \\
  $n_{b\text{-jet}} $ & 0 & 0 \\ 
 $n_\text{OSSF} $ & $\geq 1$ & $\geq 2$ \\ 
 $m_{\ell\ell} $ & $\geq 105$ GeV & $\geq 12$ GeV \\ 
 $p_T^\text{miss}$ & $\geq 200$ GeV & $\geq 200$ GeV \\ 
 $m_T$ & $\geq 160$ GeV & - \\ 
 $n_\text{exp.}$ & $2.5\pm 0.8$ & $0.97\pm 0.32$ \\ 
 $n_\text{obs.}$ & 0 & 0 \\ [1ex] 
 \hline
\end{tabular}
\caption{Relevant multilepton signal regions, taken from ref. \cite{Sirunyan:2017lae}. If the leading lepton is muon, while the rest of the leptons are electrons, the leading muon $p_T$ has to be greater than 25 GeV for both of signal regions.}
\label{tab:cuts}
\end{table}

Again we compare the official ATLAS same-sign dilepton bounds against the ones obtained from recasting the CMS multilepton search. Similar to the previous case, the multilepton signature mainly comes from $H^{++}H^-$ production. For $v_t = 10^{-9}$ GeV, the singly-charged Higgs decays into a lepton and a neutrino, while the doubly-charged Higgs decays into same-sign dilepton pair. This gives rise to $3\ell+\nu$ final state. The most significant multilepton signal region is A44,  defined in table~\ref{tab:cuts}, which requires three leptons and a large missing energy. 
We obtain the bound $m_{H^{++}}\gtrsim 740$ GeV, which is stronger than the ATLAS same-sign dilepton bound. 
The reason for a stronger bound is because the signal region considered contains little SM background. 
However, it is possible that the ATLAS same-sign dilepton bound can be comparable to, or even stronger than, the multilepton bounds when all the possible production and decay channels are combined.

The effect of other decay channels in constraining the model parameter space becomes ubiquitous as $v_t$ increases.  As example, let us consider the case of $v_t=10^{-7}$ GeV. For quite large mass splitting, i.e., $\Delta m \gtrsim20$ GeV, the $H^+$ decays predominantly into $H^{++} W^{-}$. Consequently, the final state for $H^{++}H^{-}$ production channel contains four leptons and one $W$ boson. In contrast to the case of $v_t=10^{-9}$ GeV, the stronger bounds are obtained from the signal region G05, which involves at least four leptons and large missing energy. Since the expected background in the G05 signal region is lower than the A44 signal region, the bound for  $v_t = 10^{-7}$ GeV is slightly stronger than the bound for $v_t = 10^{-9}$ GeV, as depicted in figure~\ref{fig:allbounds}. But, when mass splitting becomes less than 20 GeV, the channel $H^+\to \ell^+ \nu$ becomes sizable, so again, A44 is the most constraining signal region. That explains why in this region the bound is weaker than that of $10^{-9}$ GeV.

For intermediate value of $v_t$ (i.e., $v_t\sim 10^{-4}$ GeV), one can see from figure~\ref{fig:hpvtdm200} 
that, for the majority of the parameter space (with $\Delta m > 5$ GeV), $H^+$ decays mainly into $H^{++}W^{-*}$, while $H^{++}$ can decay into the same-sign diboson and dilepton with roughly the same rates. For sufficiently large $v_t$, where the decay of $H^{++}$ is dominated by $W^+W^+$, it leads to a final state with five $W$ bosons. On the other hand, if $H^{++}\to \ell^+\ell^+$ dominates, one gets a final state with one $W$ and four leptons. Since the leptonic branching fraction of $W$ is about $20\%$, we expect that in both scenarios the G05 signal region of CMS multilepton search~\cite{Sirunyan:2017lae} will produce the strongest bound. We can see from figure~\ref{fig:allbounds} that $m_{H^{++}}\gtrsim 470$~GeV for $v_t=10^{-4}$ GeV and $\Delta m\gtrsim 5$ GeV, which is slightly stronger than ATLAS bound for the same triplet vev value. But when the mass splitting is less than 5 GeV, the bound gets weaker because now $\ell^+\nu$ starts to dominate $H^+$ decay, leading to $3\ell+\nu$ final state. We have known from the previous discussion that such final state, consistent with A44 signal region, has a larger SM background, and thus, explaining the relaxed bound. 

\subsection{$\Delta m < 0$} \label{sec:deltamnegative}
For $\Delta m < 0$, the most dominant scalar pair production channels are the $HA$ and the $H^+H/A$. The $H^{++}H^-$ and $H^{++}H^{--}$ production cross-sections, however, remain roughly the same as in the $\Delta m>0$ case for a fixed value of $m_{H^{++}}$. We will see that we need to utilize all of these production channels to constrain the type-II seesaw model.

Let us first consider the case where $v_t$ is small. The $H$ and $A$ scalars decay predominantly into neutrinos, rendering them invisible to the detector. At the same time, the dominant decay modes for $H^+$ and $H^{++}$, for sufficiently small $v_t$, are lepton pairs. Thus, for sufficiently small $v_t$, we expect to get a similar bound to the ones obtained from the $H^{++}H^-$ and the $H^{++}H^{--}$ production channels in the $\Delta m>0$ scenario. This is the case for $v_t=10^{-9}$ GeV and $v_t=10^{-7}$ GeV with $\Delta m  \gtrsim -20$ GeV. For $v_t=10^{-7}$ GeV with $\Delta m \lesssim -20$ GeV, the $H^{++}$ decays into $H^+$ and $W^+$ and the $H^+$ decays into $W^+$ and $H$/$A$. As has already been noted, both $H$ and  $A$ decay invisibly into neutrinos. Therefore, we do not expect any LHC constraint in this particular case. However, if $H$ and $A$ masses are smaller than $m_h/2$, then the decays of $h \rightarrow H H, AA$ are open. These decay modes are constrained by the invisible decay width of the $h$~\cite{Khachatryan:2016whc, Aad:2015txa, Aaboud:2018sfi, Sirunyan:2018owy}.\footnote{The bound from invisible Higgs decay is stronger than the monojet bounds~\cite{ATLAS:2017dnw, Aaboud:2017phn}, which we found to be $m_{H/A} > 40$ GeV.} 

Both $H$ and $A$ start to have substantial visible decays for $v_t \gtrsim 10^{-4}$ GeV. If their masses are large enough, the decay of the $H$ will be dominated by the gauge boson pair or the $hh$ pair, while the dominant decay mode of $A$ will be to $Zh$. Moreover, the $H^+$ decays into $W^+H/A$ for most of the parameter space. In this case, the most important production channel is the pair production $H^+H/A$.  This production channel leads to final states containing some combination of $W$, $Z$ and $h$. Since $W$, $Z$ and $h$ all have sizable decay into leptons, the LHC multilepton search is the most sensitive channel in this case. We find that the signal region G05 is the most relevant signal region in this case.

When the mass of $H$ is below the $W^+W^-$ threshold, it decays mainly into $b \bar b$. Similarly, the pseudoscalar $A$ also decays primarily into a $b$-quark pair since its mass is below the $hZ$ threshold. Therefore, the LHC multilepton search does not apply. 
There are some relevant LHC searches in this case. For example the CMS search for pair-produced resonances decaying into quark pairs~\cite{Sirunyan:2018rlj} is relevant for constraining the $H A$ pair production. However, the QCD backgrounds for this process completely overwhelm the weak scale production of the $H A$ signal. 
To avoid the large QCD background, we instead consider the LEP MSSM Higgs boson search~\cite{Schael:2006cr}.
The LEP searched for the $H A$ pair production at the $e^+ e^-$ collider with the center of mass energy up to 209 GeV. 
One of the scenarios presented in ref.~\cite{Schael:2006cr} is to assume that both $H$ and $A$ decay 100\% into $b \bar b$ pair. There, the bounds are presented as the ratio of the maximum allowed $H A$ production cross-section to the $hZ$ production cross-section multiplied by kinematic factors. Hence, after taking into account the branching fraction of $H$ and $A$ into $b \bar b$, we can compare directly the bounds presented in ref.~\cite{Schael:2006cr} with the cross-section in our model. The bound is determined to be $m_{H} > 97.5$ GeV from the LEP search.

\begin{figure}
       \centering
                \includegraphics[width= 0.6\textwidth]{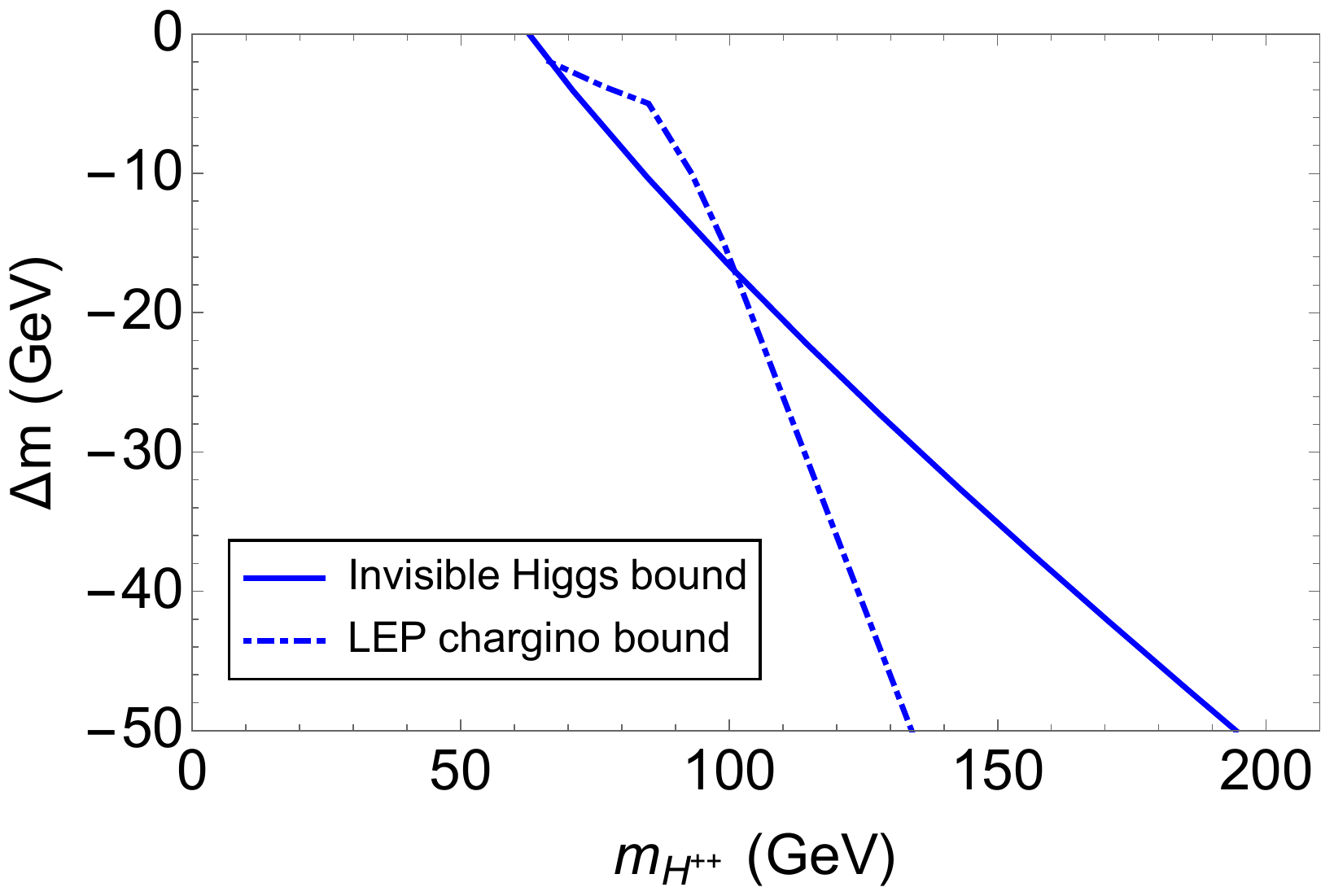}
        \caption{The bounds for $v_t = 10^{-4}$ GeV from invisible Higgs branching fraction \cite{Khachatryan:2016whc, Aad:2015txa, Aaboud:2018sfi, Sirunyan:2018owy} and the LEP chargino searches \cite{Abbiendi:1999ar}. The region on the left of each line is excluded.}
         \label{fig:invisiblevschargino}
\end{figure} 

For $v_t = 10^{-4}$ GeV and relatively light neutral scalars, both $H$ and $A$ decay mainly into neutrinos. Hence the bound on invisible decay $m_{H/A} > 62.5$ GeV applies in this case, while the LEP MSSM Higgs boson search is irrelevant. However, in this case the LEP search for chargino~\cite{Abbiendi:1999ar, Abbiendi:2003sc} can be used to constrain the model parameter space. Due to the cuts in the analysis, we find that the most relevant LEP search is the OPAL chargino search at $\sqrt{s} = 189$ GeV using 182.1 pb$^{-1}$ of data~\cite{Abbiendi:1999ar}. In their analysis, the OPAL collaboration look for a pair production of charginos followed by the chargino decays into a $W$ and a neutralino. In our model, we have a pair production of $H^+H^-$ followed by the $H^\pm$ decay into $W^\pm$ and $H$/$A$. Since both $H$ and $A$ decay invisibly, the $H^+H^-$ pair production mimics the charginos pair production. We find that for a large enough $\Delta m$, the bound obtained from the chargino search is $m_{H^+} > 83$ GeV. This bound is better than the invisible Higgs decay bound, $m_{H} > 62.5$~GeV, only for $\Delta m > - 17$ GeV. For smaller mass splitting values, the bound becomes worse. 
The comparison between the two bounds are shown in figure \ref{fig:invisiblevschargino}.

As mentioned above, collider bounds depend strongly on the value of $v_t$ and $\Delta m$ since they determine the branching fraction of each decay channels. The bounds also depend on the value of $\sin\alpha$, albeit not as strong as the previous two parameters. In figure \ref{fig:allbounds}, we show the bounds for various values of $v_t$. For a particular value of $v_t$ and $\Delta m$, we varied $\sin\alpha$ to be between $\pm 24.6v_t/v$ and took the weakest bounds. Additionally we plot the bounds from the electroweak precision test discussed in section~\ref{sec:ewpc} and the bounds from the invisible Higgs decay; both  are independent of $v_t$ and $\sin\alpha$.

\begin{figure}
       \centering
                \subfloat[$v_t = 0.1$ GeV]{\includegraphics[width= 0.49\textwidth]{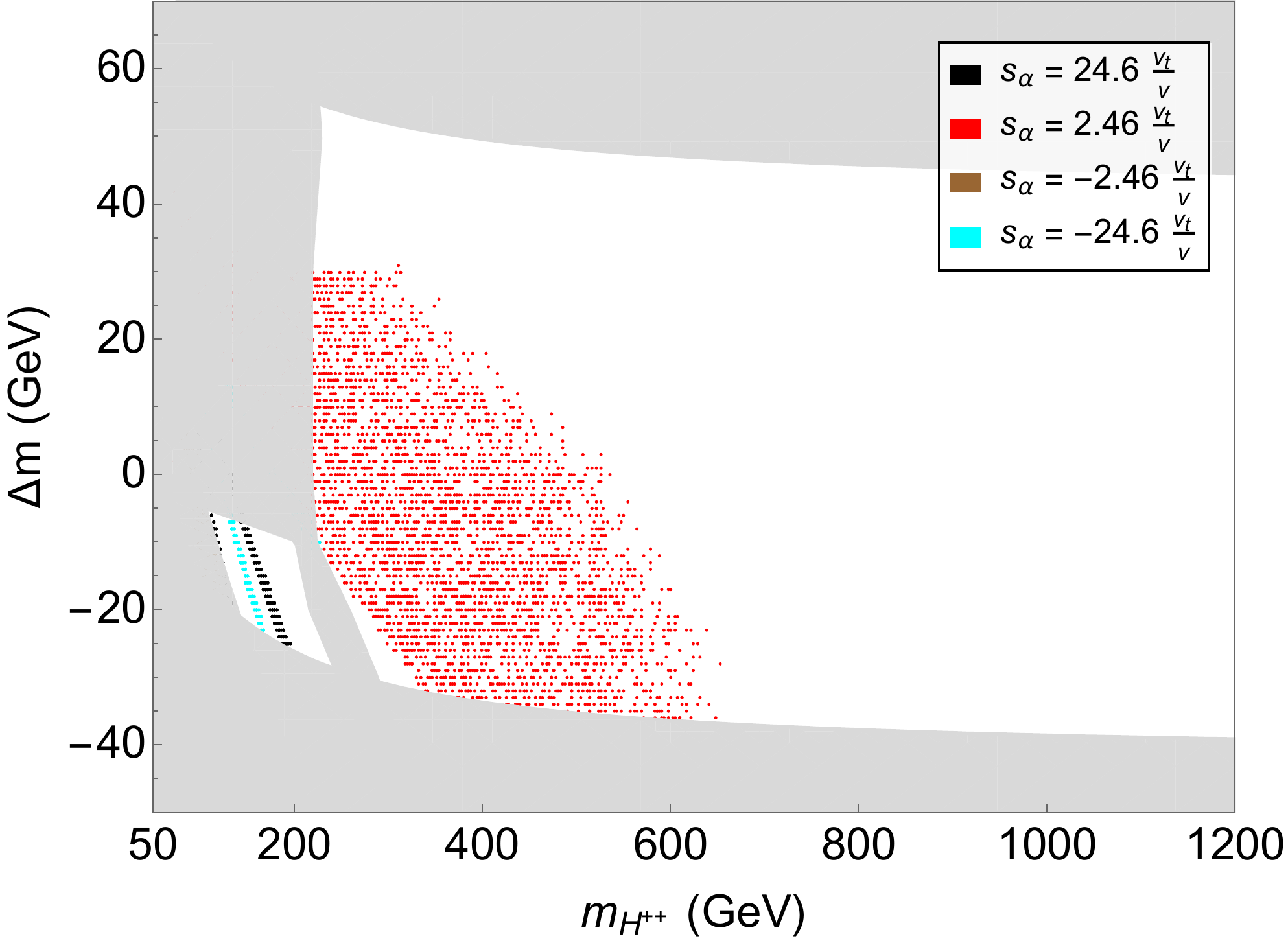}\label{fig:vt1}}  \,\,\,
        \subfloat[$v_t = 10^{-4}$ GeV]{\includegraphics[width= 0.49\textwidth]{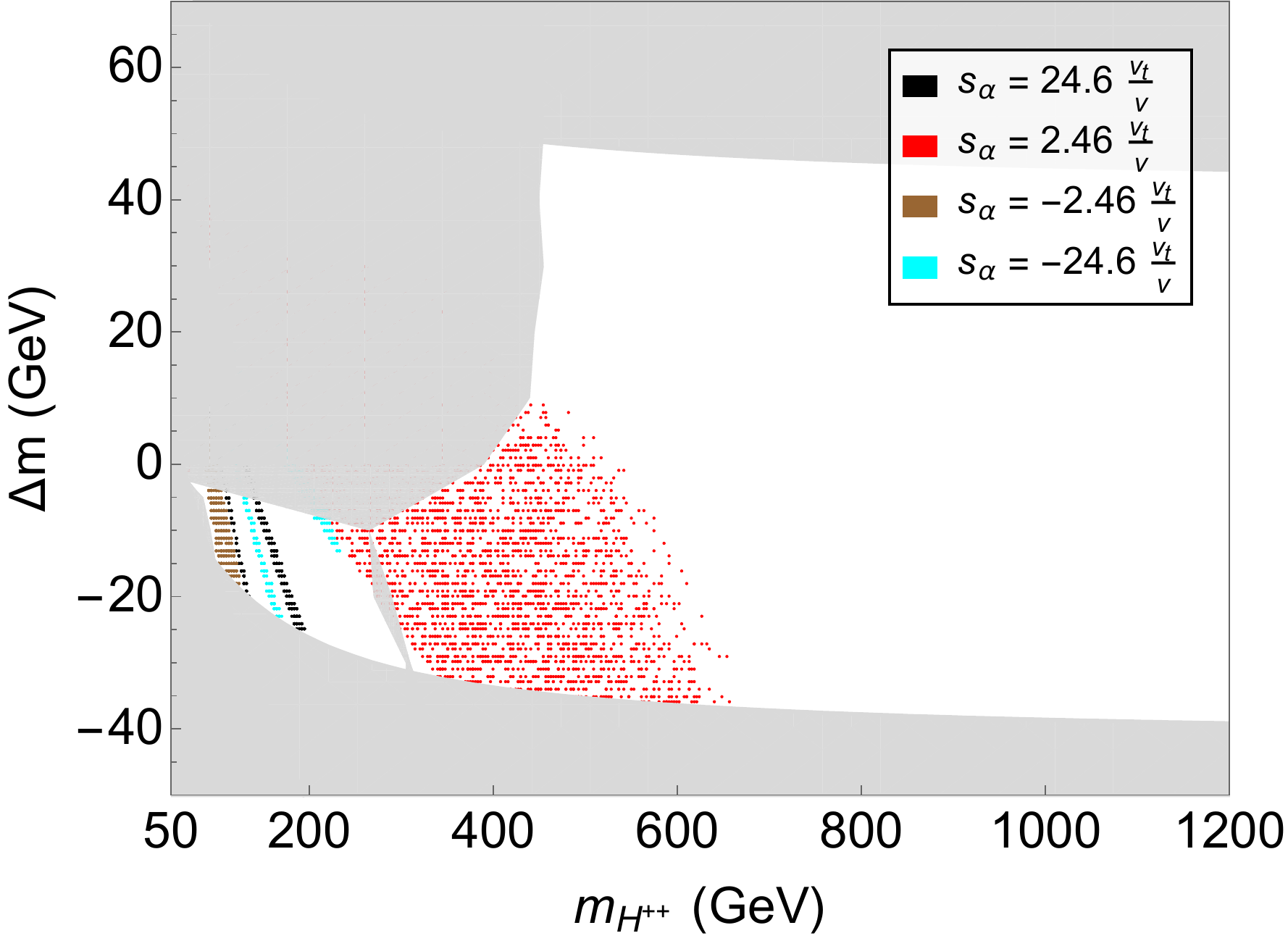}\label{fig:vt4}}  \\
                        \subfloat[$v_t = 10^{-7}$ GeV]{\includegraphics[width= 0.49\textwidth]{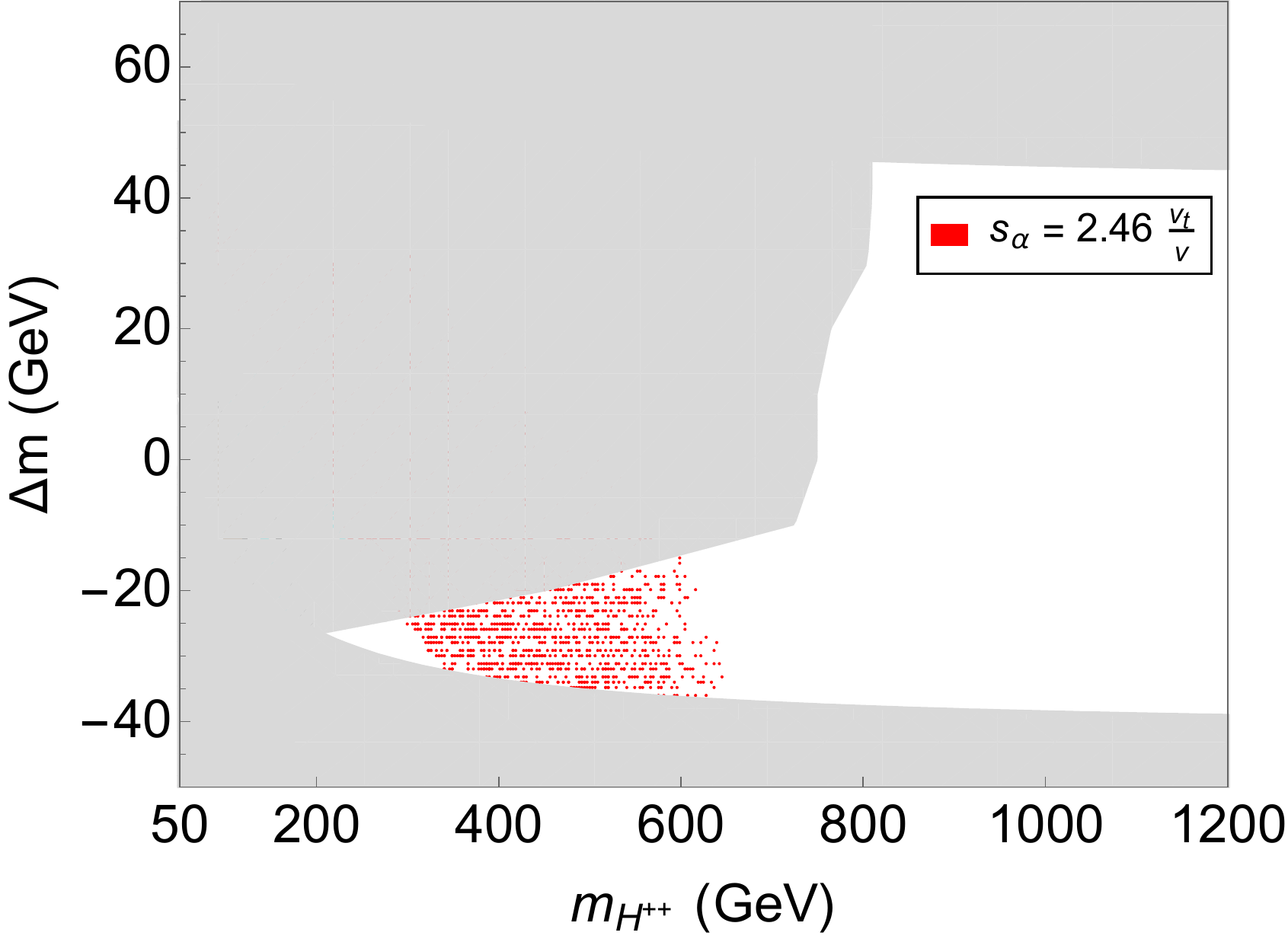}\label{fig:vt7}}  \,\,\,
                        \subfloat[$v_t = 10^{-9}$ GeV]{\includegraphics[width= 0.49\textwidth]{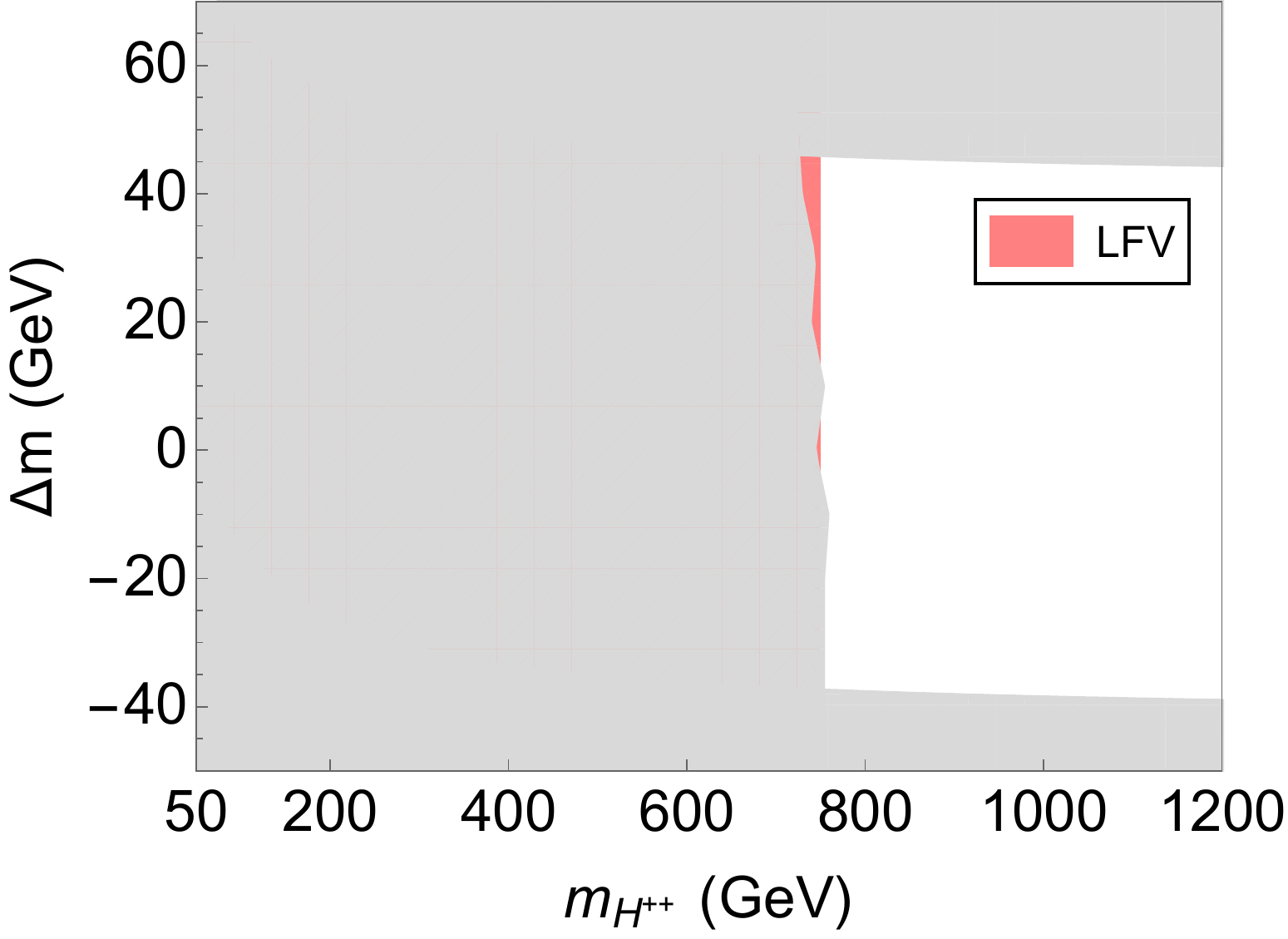}\label{fig:vt9}} 

        \caption{The combination of allowed regions from the Higgs data, shown from figure \ref{fig:higgsdata}, and the exclusion regions of colliders, Higgs invisible searches and EWPT, shown in figure~\ref{fig:allbounds}. In the figure, we take $v_t = 0.1$ GeV, $10^{-4}$ GeV, $10^{-7}$ GeV and $10^{-9}$ GeV. Additionally for each value of $v_t$, we take the benchmark cases $s_\alpha = \pm 2.46 v/v_t$ and $\pm 24.6 v/v_t$ for the allowed regions. The gray shaded region is the combination of exclusion regions of colliders, Higgs invisible searches and EWPT. }
         \label{fig:sinalpha}
\end{figure}

From the plot we can see that for the case of $\Delta m > 0$, LHC can constrain the parameter space of the model very well. The bounds in this case are $m_{H^{++}} \gtrsim 200$ GeV, with the weakest bound comes from a large value of $v_t$, in which the $H^{++}$ decays into $W^+W^+$. Hence the search from this channel with a further addition of $H^{++}H^{-} \rightarrow W^{+}W^{+}W^{-}$ channel will push the constraint for the case of $\Delta m > 0$.

In the case of $\Delta m < 0$ and  $v_t \gtrsim 10^{-4}$ GeV, there is no current collider bound when the mass of $H$ falls below the $W^+W^-$ mass threshold. In this case, both $H$ and $A$ decay into $b\bar b$ or $\nu \nu$, depending on the value of $v_t$. The two decay channels are difficult to probe because of large backgrounds in the former case or missing energy in the final states in the latter. As a result, the value of $m_H$ (or $m_A$) can be as low as $m_h/2$, which is nothing but the invisible Higgs decay bound. Likewise, in the case where all the scalar masses are almost degenerate, the mass of ${H^{++}}$ can also be close to $m_h/2$ because the decay of $H^{++}$ will not proceed through very visible decay channels unlike in the case of $\Delta m > 0$. The final states in this case, after subsequent cascade decays, will be very soft leptons and $H/A$; both are indeed very difficult to probe at the LHC.

The bounds that are presented in figure~\ref{fig:allbounds} are {\it general} in the sense that they admit all possible values of $\sin\alpha$ for each $v_t$ considered. It is interesting to see  what will happen if these bounds are overlaid on the scattered plots of figure~\ref{fig:higgsdata}, which are obtained by fixing $\sin\alpha$ and $v_t$. As can be seen in figure~\ref{fig:sinalpha}, collider bounds can eliminate quite significant area, especially for low $v_t$; this is evident in the case of $v_t=10^{-7}$ GeV and $v_t=10^{-9}$ GeV. None of the four benchmark points being discussed can survive in the latter, while in the former, only the case of $s_\alpha=2.46 v_t/v$ can barely escape the exclusion. As a matter of comparison, in that figure, we also show the LFV bound for $v_t=10^{-9}$ GeV, that is $m_{H^{++}}\gtrsim 750$ GeV. As discussed in section~\ref{sec:LFV}, this bound is obtained by assuming normal neutrino mass hierarchy with $m_1=0$. For a higher value of $v_t$ the LFV bounds are less relevant.

\section{Conclusions and discussions}
\label{sec:conclusion}
The type-II seesaw mechanism is one of the simplest framework to incorporate neutrino masses into the SM. Despite its simplicity, the type-II seesaw model offers rich phenomenology. In our work, we have investigated various constraints on the type-II seesaw model parameter space, from the low energy observables constraints to the theoretical bounds to the collider constraints. 

Constraints from low energy observables can greatly reduce the size of the viable parameter space. The electroweak $\rho$ parameter constrains the triplet vev to $v_t\lesssim 4.8$ GeV. This constraint ensures the spectrum of the scalar bosons satisfies an approximate relation $m_{A}^2 \simeq \sin^2\alpha\,m_h^2 + \cos^2\alpha\,m_{H}^2 \simeq 2 m_{H^{+}}^2 - m_{H^{++}}^2$. As a result we can parametrize the parameter space of the model by three parameters: the CP-even mixing angle $\alpha$, the mass splitting $\Delta m$ and the doubly-charged mass $m_{H^{++}}$ ($m_h$ is taken to be 125 GeV). The electroweak $S$ and $T$ parameters further reduce the allowed parameter space to $-40\text{ GeV}\le\Delta m\le 50$ GeV. 
Constraints from the lepton flavor violation processes are also relevant for low values of $v_t$. These processes become the most constraining observables in the case of $v_t = 10^{-9}$ GeV. They place the bounds on the doubly-charged Higgs mass $m_{H^{++}} \gtrsim 750$ GeV for the NH case with $m_1 = 0$ . The bounds could be stronger for other neutrino mass scenarios.

Theoretical and collider bounds further constrain the type-II seesaw model parameter space. Their contributions are intertwined. From the 125-GeV Higgs boson properties, one expects the mixing angle $\alpha$ to be small. Moreover, the invisible branching ratio of the 125-GeV Higgs boson places a strong constraint on the neutral scalar masses, $m_H,m_A\ge m_h/2$. Additionally, perturbativity of the scalar quartic couplings further dictates that $v_t$ and $\sin\alpha$ are correlated, as can be seen in figure~\ref{fig:savt}.
This, coupled with the smallness of $v_t$ and $\sin\alpha$, further simplifies the relations among the scalar masses: $m_{A}^2 \simeq m_{H}^2 \simeq 2 m_{H^{+}}^2 - m_{H^{++}}^2$. Finally, stability of the electroweak vacuum, discussed in section~\ref{sec:theoretical}, implies that, for a given value of $\Delta m$, $m_{H^{++}}$ in most of parameter space cannot be arbitrarily large. This is reflected, for example, as a boundary on the right of the viable parameter space shown  in figure~\ref{fig:higgsdata}.    

Having identified the viable parameter space, we can further probe them with collider searches for the extra scalar bosons: $H$, $A$, $H^+$ and $H^{++}$.
The collider signatures of these scalars depend heavily on the mass splitting $\Delta m$ and $v_t$. For the case of $\Delta m > 0$, the lightest new scalar is $H^{++}$. There exist official CMS and ATLAS results from the $H^{++}H^{--}$ production channel. They considered the $W^+W^+$ and $\ell^+\ell^+$, which are the main decay products of the $H^{++}$ in the cases of high and low $v_t$, respectively. We compare these official bounds with the bounds from the recasted multilepton signals from the $H^{\pm}H^{\mp\mp}$ and the $H^{\pm\pm}H^{\mp\mp}$  production channels. We find that the multilepton bounds are comparable to or often better than the official LHC bounds.

The part of parameter space with $\Delta m<0$ is not covered by any official LHC searches. We can again use the multilepton search to probe this region provided that the the heavy neutral scalars can decay into $W^+W^-$ or $hh$. However, when $H$ and $A$ are too light, both of them decay into $b \bar b$ or $\nu \nu$, depending on the value of $v_t$. In this case, the multilepton search becomes irrelevant. Hence we believe this region of parameter space has not been constrained by any of the LHC searches. There are several possible signatures that might be able to probe this particular region of the parameter space. If the branching fraction of the $H/A$ decay to $b \bar b$ are significant, the final states of $H^{\pm} H/A$ production channels are $W^* b \bar b b \bar b$. This, in turn, leads to a signal with four $b$-jets, a lepton and missing energy. Since the lepton and the missing energy in this case come from the leptonic decay of the off-shell $W$, the missing energy requirement should be much lower than the ones used in supersymmetry searches. Additionally, if we consider the $H^+H^-$ production channel, we can have four $b$-jets, two leptons and a missing energy signal from the cascade decay of $H^+$ and $H^-$. We leave the LHC sensitivity to these channels for possible future works. The summary of the bounds are shown in figures \ref{fig:allbounds} and \ref{fig:sinalpha}. 

\begin{acknowledgments}
The work of RP is supported by the Parahyangan Catholic University under grant no. III/LPPM/2019-01/42-8. RP thanks NCTS, Hsinchu and ICTP, Trieste for the hospitality during part of this work was carried out. He also benefited from the use of the ICTP Argo cluster. 
JJ acknowledges the support he receives from the Indonesian Institute of Sciences under grant no.~B-288/IPT/HK.02/II/2019.
The work of PU has been supported in part by the Thailand Research Fund under contract no.~MRG6280186, and the Faculty of Science, Srinakharinwirot University under grant no.~222/2562. We also thank Tim Tait and Konstantin Matchev for organizing TASI 2011, a school that kicks start our fruitful theoretical particle physics collaboration in Southeast Asia.

\end{acknowledgments}

\appendix

\section{Dependence on neutrino mass parameters} \label{sec:numass}

\begin{table}[t!]
\centering
\begin{tabular}{||c | c | c | c | c | c | c | c ||} 
 \hline
   & NH & BP1 & IH & BP2 & QD& QD & BP3\\
    &  &   &  &  & (normal) & (inverted) & \\ [0.5ex]
 \hline\hline
 $e^+ e^+$ & 1\% & 0\% & 47\% & 50\% & 30\% & 33\% & 33\%\\ 
 $\mu^+ \mu^+$ & 38\% & 30\% & 8\% & 12\% & 35\% & 30\% & 33\%\\ 
 $\tau^+ \tau^+$ & 21\% & 30\% & 17\% & 12\% & 33\% & 32\% & 33\%\\ 
 $e^+ \mu^+$ & 4\% & 1\% & 1\% & 0\% & 1\% & 3\% & 0\%\\ 
  $e^+ \tau^+$ & 1\% & 1\% &  1\% & 0\% & 1\% & 2\% & 0\%\\ 
   $\mu^+ \tau^+$ & 35\% &38\% & 26\% & 25\% & 0\% & 0\% & 0\%\\ 
 \hline
  Bounds (GeV) & 745 & 723 & 705 & 716 & 755 & 755 & 761 \\
  \hline
\end{tabular}
\caption{The columns NH, IH and the two QD are the branching ratios of $H^{++}$ for $v_t = 10^{-9}$ GeV. These values also hold for $v_t = 10^{-7}$ GeV and $\Delta m \gtrsim -20$ GeV. For smaller values $\Delta m$, the doubly-charged Higgs decays mainly into $W^{*+} H^+$. The bounds shown are for $\Delta m = 0$. The columns BP1, BP2 and BP3 are the benchmark points in the CMS analysis \cite{CMS:2017pet}.}
\label{tab:bfhpp}
\end{table}

In this appendix,
we consider the effect of varying the PMNS parameters and the neutrino masses on the LHC bounds. We will consider three cases: (1) the normal hierarchy with massless lightest eigenstates (NH), this will be our benchmark case; (2) The inverted hierarchy with massless lightest eigenstates (IH); and (3) The  the lightest neutrino eigenstates is having a mass of 0.1 eV. In case (3), for a low enough $v_t$, the $H^{++}$ decays almost equally into each flavor of dilepton pair for both the normal and the inverted hierarchy scenarios. Therefore the collider bounds for case (3) will be practically the same for the two mass hierarchy scenarios. We refer to case (3) as the quasi-degenerate (QD) scenario. In all three cases above we take the Majorana phases to be zero. The branching ratio for the doubly-charged Higgs in the three cases are given in table~\ref{tab:bfhpp}. 

\begin{figure}
       \centering
                        \subfloat[$v_t = 10^{-7}$ GeV]{\includegraphics[width= 0.49\textwidth]{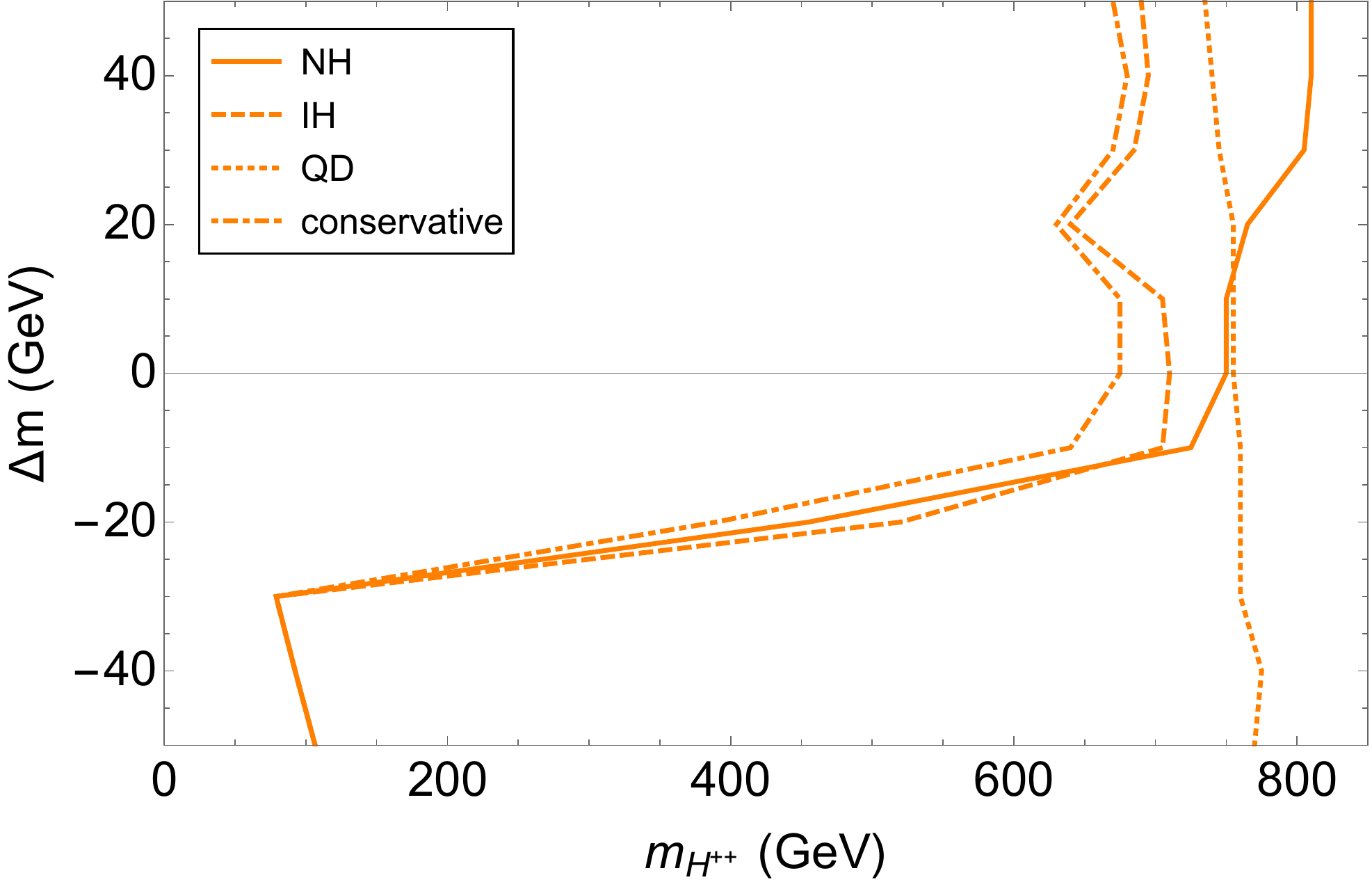}\label{fig:vt1em7varneu}} \,\,\,
                        \subfloat[$v_t = 10^{-9}$ GeV]{\includegraphics[width= 0.49\textwidth]{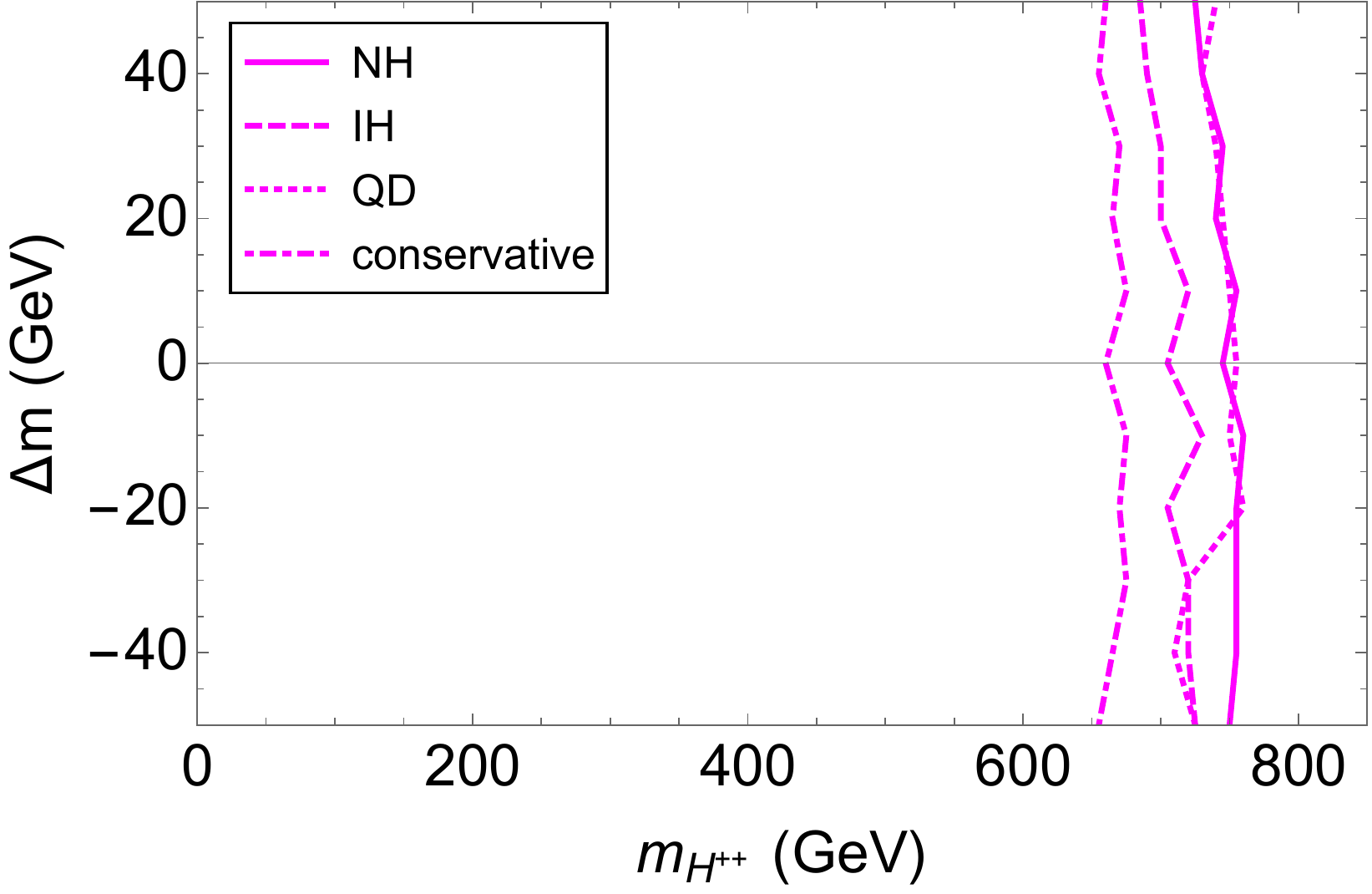}\label{fig:vt1em9varneu}} 

        \caption{The bounds for various neutrino mass cases. The region on the left of each line is excluded.}
         \label{fig:boundvarneu}
\end{figure} 

In the case of $v_t \lesssim 10^{-4}$ GeV, the triplet Yukawa couplings are small enough. Hence the variation on the neutrino mass matrix does not change the collider bounds significantly. The collider bounds for $v_t =10^{-7}$ GeV and $10^{-9}$ GeV are shown in figure~\ref{fig:boundvarneu}. For $v_t = 10^{-9}$ GeV, the collider bounds for all three cases are approximately $m_{H^{++}}\gtrsim 750$ GeV. Note that the NH and QD bounds are about the same and are slightly stronger than the IH case. This is because, in the NH and the QD cases, the $H^+$ and the $H^{++}$ have significant branching fractions into muons, which can be cleanly detected. In the case of $v_t = 10^{-7}$ GeV, the bounds for all three cases are also approximately the same for $\Delta m \ge -20$ GeV. 
For $\Delta m \le -20$ GeV, the collider bounds for the NH and the IH cases get weaker because they are in the region of parameter space that can only be constrained by the invisible Higgs decay width (which happens only when $m_{H,A} < m_h/2$). For the QD case, the triplet Yukawa couplings are large enough so that the dilepton channels are still dominant. Hence collider bound remains practically the same.

The bounds obtained from the multilepton channel are in general better than our recasted ATLAS search for doubly-charged Higgs. The reason is that the ATLAS search only considers $H^{++}H^{--}$ production mechanism, while we also include the $H^{\pm\pm}H^{\mp}$ production channel in deriving our multilepton bound. Moreover, we only consider one decay channel when recasting the bound. There exists a CMS analysis for the singly- and doubly-charged Higgs \cite{CMS:2017pet} using only 12.9 fb$^{-1}$ of data at 13 TeV. In their analysis, CMS show that the inclusion of the $H^{\pm\pm}H^{\mp}$ production channel improves the bound on $m_{H^{++}}$. However, when considering the $H^{\pm\pm}H^{\mp}$ production channel, CMS assume that the $H^+$ decays only into $\ell \nu$. Moreover they also assumed that the singly- and doubly-charged Higgs masses are degenerate. This corresponds to the $v_t = 10^{-9}$ GeV and $\Delta m = 0$ case in our benchmark scenario. CMS also consider four more scenarios of the $H^{++}$ branching ratios. Three of them are relatively similar, while not exactly the same, with our cases. The CMS benchmark point BP1 is similar to our NH case, BP2 is comparable with our IH scenario, and BP3 is close to our QD case. Hence for illustrative purpose, we compare the bounds for these benchmark points.  Since CMS can optimize the cuts for their particular search purpose, their bounds are comparable to our recasted multilepton bounds, despite using only a third of collision data.

As mentioned above, the CMS analysis assume the singly-charged Higgs decay into $\ell \nu$, which is true for the case of $v_t = 10^{-9}$ GeV. However, this assumption is not necessarily valid in other parts of parameter space. For example, in the case where $v_t = 10^{-7}$ and $\Delta m \gtrsim 10$ GeV, the $H^+$ decays mostly to $W^{-} H^{++}$, resulting in a $3\ell + \nu$ final state. Therefore we suggest that, besides considering the decay of $\ell \nu$ for the singly-charged Higgs, CMS should also include the $H^{+}\to W^{(*)-} H^{++}$ decay mode in their future analysis. In other words, CMS should consider relaxing degenerate triplet mass assumption.

\begin{table}[t!]
\centering
\begin{tabular}{||c | c ||} 
 \hline
Decay channel & Branching ratio (\%)  \\
\hline
 $e^+ e^+$ & 4\%\\ 
 $\mu^+ \mu^+$ & 7\%\\ 
 $\tau^+ \tau^+$ & 18\%\\ 
 $e^+ \mu^+$ & 3\%\\ 
  $e^+ \tau^+$ & 3\%\\ 
   $\mu^+ \tau^+$ & 65\% \\ 
 \hline
\end{tabular}
\caption{The branching fractions of $H^{++}$ for the most conservative scenario. The mass hierarchy used to produce the branching ratio is the normal hierarchy with $\sin^2\theta_{23} = 0.428$, $m_1= 0.0106$ eV,  $\alpha_1 = 23.4^\circ$ and
$\alpha_2 = 194.7^\circ$. For other PMNS parameters, we use parameters shown in the table~\ref{nu-fit}.}
\label{tab:worst}
\end{table}

Departing from the benchmark cases, we investigate how the variation in each PMNS parameter impacts the derived collider bounds. We find that there are four parameters that can significantly change the branching ratio of $H^{++}$: $\theta_{23}$, the lightest neutrino mass and the Majorana phases $\alpha_1$ and $\alpha_2$. Varying the other PNMS parameters within their allowed $3\sigma$ range, as given in ref.~\cite{Esteban:2018azc}, only affect the $H^{++}$ branching ratios by at most 2\%. Therefore, we vary the four parameters mentioned above to obtain the most conservative (weakest) bound. We find that the weakest bound is achieved in the normal hierarchy scenario with the $H^{++}$ branching fractions shown in table~\ref{tab:worst}. This conservative bound is obtained when the sum of the branching fractions $H^{++}$ to $e^+e^+$, $\mu^+\mu^+$ and $e^+\mu^+$ is at its minimum. The conservative bounds are shown in figure~\ref{fig:boundvarneu} and labeled as conservative.

Finally, we can compare our collider bounds to the LFV bounds. 
We use LFV bounds discussed in section~\ref{sec:LFV} where some dependence on neutrino mass hierarchies are apparent. 
One can see in the case of $v_t = 10^{-9}$ GeV that LFV processes in general induce stronger bounds, except for the NH case where the LFV bound is comparable to the collider bound.  As discussed in section~\ref{sec:LFV}, the LFV bounds will get weaker by a factor of $(v_t/10^{-9}~\rm GeV)$ as $v_t$ increases. So, already for $v_t=10^{-7}$ GeV, the strongest LFV bound, achieved in the IH case, is 730 GeV. This is more or less the same as the collider bound for $\Delta m\gtrsim -20$ GeV. 

\bibliography{reference} \bibliographystyle{JHEP}

\end{document}